\documentclass[a4paper,fleqn,usenatbib]{mnras}

\usepackage{lscape}
\usepackage{graphicx}
\usepackage{amssymb}
\usepackage{amsmath}
\usepackage{url}
\usepackage{bm}
\usepackage{aas_macros}
\usepackage{rotating}

\usepackage{multirow}
\usepackage{color} 

\title[]{The VMC survey - XLVII. Turbulence-Controlled Hierarchical Star Formation in the Large Magellanic Cloud}

\author[A.~E.~Miller et al.]{Amy E.~Miller,${^{1,2}}$\thanks{E-mail:
  amiller@aip.de (AEM)} Maria-Rosa L. Cioni,$^{1}$ Richard de Grijs,$^{3,4}$
  Ning-Chen Sun,$^5$
  \newauthor
  Cameron P. M. Bell,$^{1}$ Samyaday Choudhury,$^{3,4}$ Valentin D. Ivanov,$^6$
  \newauthor
  Marcella Marconi,$^7$ Joana Oliveira,$^8$ Monika Petr--Gotzens,$^6$ Vincenzo Ripepi,$^7$ and
  \newauthor
  Jacco Th. van Loon$^8$\\
  $^1$Leibniz-Institut f{\"u}r Astrophysik Potsdam, An der Sternwarte 16, D-14482 Potsdam, Germany\\
  $^2$Institut f\"{u}r Physik und Astronomie, Universit\"{a}t Potsdam, Haus 28, Karl-Liebknecht-Str. 24/25, D-14476 Golm (Potsdam), Germany\\
  $^3$School of Mathematical and Physical Sciences, Macquarie University, Balaclava Road, Sydney, NSW 2109, Australia\\
  $^4$Research Centre for Astronomy, Astrophysics and Astrophotonics, Macquarie University, Balaclava Road, Sydney, NSW 2109, Australia\\
  $^5$Department of Physics and Astronomy, University of Sheffield, Hicks Building, Hounsfield Road, Sheffield S3 7RH, UK\\
  $^6$European Southern Observatory, Karl-Schwarzschild-Str. 2, D-85748 Garching bei M\"{u}nchen, Germany\\
  $^7$INAF -- Osservatorio Astronomico di Capodimonte, Salita Moiariello 16, I--80131 Naples, Italy\\
  $^8$Lennard-Jones Laboratories, Keele University, ST5 5BG, UK\\
  }

\begin{document}

\date{Accepted ?, Received ?; in original form ?}

\pagerange{\pageref{firstpage}--\pageref{lastpage}} \pubyear{2021}

\maketitle

\label{firstpage}

\begin{abstract}
We perform a statistical clustering analysis of upper main-sequence stars in the Large Magellanic Cloud (LMC) using data from the Visible and Infrared Survey Telescope for Astronomy survey of the Magellanic Clouds. We map over 2500 young stellar structures at 15 significance levels across $\sim$120 square degrees centred on the LMC. The structures have sizes ranging from a few parsecs to over 1 kpc. We find that the young structures follow power-law size and mass distributions. From the perimeter--area relation, we derive a perimeter--area dimension of 1.44$\pm$0.20. From the mass--size relation and the size distribution, we derive two-dimensional fractal dimensions of 1.50$\pm$0.10 and 1.61$\pm$0.20, respectively. We find that the surface density distribution is well-represented by a lognormal distribution. We apply the Larson relation to estimate the velocity dispersions and crossing times of these structures.
Our results indicate that the fractal nature of the young stellar structures has been inherited from the gas clouds from which they form and that this architecture is generated by supersonic turbulence. Our results also suggest that star formation in the LMC is scale-free from 10 pc to 700 pc.
\end{abstract}


\section{Introduction}
\label{introduction}

Stars are observed in galaxies both as individual field stars and grouped in clusters, associations and complexes. Young stars with ages less than 100 Myr are more spatially clustered than evolved, older stars. These spatially clustered young star groups are hierarchically organised: larger, lower-density groups contain one or more smaller, higher-density groups, which then fragment into even smaller and more compact groups \citep{elmegreen2010}. This hierarchical architecture is also observed for gas structures in the interstellar medium \citep[ISM;][]{elmegreen1996,blitz1999}. Observations and simulations suggest that the organisation of young stellar structures originates from the fractal distribution of the ISM \citep{elmegreen1996,efremov1998,elm&elm2001,vs2016}. This is evidence of hierarchical star formation, also known as `scale-free' or fractal star formation \citep{gomez1993,larson1995,bate1998}. 


Classically, it was thought that stars could form in either the field of a galaxy as single stars or concurrently in star clusters that formed from the same molecular cloud. This would mean that there are two distinct star formation modes: `clustered' and `field' or `distributed' star formation \citep{lada&lada2003}. However, \citet{bressert2010} found that young stellar systems exist in a continuous range of stellar densities or masses, implying that there is no bimodality in star formation and star clusters only represent the densest part of the continuum. Further investigation of the hierarchical properties of star formation is necessary to gain a better understanding of the full continuum over a large parameter range and assess whether there are any characteristic scales or distinct modes of star formation, which may correspond to some physical mechanisms. 


Hierarchical star formation has many broad-reaching astrophysical implications. For example, the universal surface brightness profiles of young massive clusters in different environments \citep{grudic2018} can be explained if clusters form through the hierarchical merging of subclusters. In addition, hierarchical mergers of subclusters may lead to a phase when the cluster becomes very compact, characterised by short dynamical time-scales, and dynamical mass segregation may occur during this very early stage \citep{allison2009}. Binary star clusters that are located very close to each other and have similar ages \citep[e.g., the Double Cluster;][]{doublecluster2001} may have formed as a consequence of hierarchical star formation and may be affected by mutual dynamical interactions. Additionally, the commonly observed multiple stellar populations in star clusters could be a consequence of hierarchical star formation: a newly formed star cluster may continue to accrete gas from its surroundings and form new stars of younger ages \citep{bekki2017}. On a larger scale, the observed ultra-compact dwarf galaxies may be the result of mergers of many star clusters in a star-forming complex \citep{urrutiazapata2019}, which is expected in hierarchical star formation. Hierarchical star formation can also explain why only a minority of core-collapse supernovae are found in star clusters \citep[e.g.][]{sun2020} while the majority are not.

The properties of the star-forming hierarchy are not yet fully understood since they have only been studied for a limited number of star-forming regions and galaxies. The goal of this paper is to better understand the star-formation hierarchy on scales between star clusters and galaxies. Performing a global study of hierarchical star formation in the Milky Way is very difficult, because we cannot view the Galaxy in its entirety and there is dust obscuration in the Galactic plane. This, therefore, turns our attention to the Milky Way’s most massive satellite galaxy, the Large Magellanic Cloud (LMC). The LMC is seen face-on and features active star formation. It is the ideal candidate to study hierarchical star formation owing to its orientation, proximity and the small line-of-sight depth \citep[ranging from 2.6 to 4.2 kpc in the disc and Bar regions;][]{subramanian2009}. The LMC exhibits signatures of interactions with its companion, the Small Magellanic Cloud (SMC), as well as with the gravitational potential and halo gas of the Milky Way. The LMC was the first nearby galaxy in which a large population of young, massive stars was observed. \citet{hodge1973} observed star clusters in the LMC and noted that their distribution is connected to the patchy distribution of H{\sc ii} regions. They also noted a pattern of star formation on 1 kpc size scales, which is related to very dense gas regions in the LMC. \citet{efremov1998} showed that the average age difference between pairs of star clusters in the LMC increases with their separation. This suggested that star formation is hierarchical in space and time. They showed that, in small regions, stars form quickly, whereas in large regions, which also contain smaller groups of stars, stars form over a longer period. \citet{bastian2009} analysed the age distributions of different stellar groups in the LMC. They found groups of younger stars to be highly substructured and nonuniform, and that this nonuniform distribution dissolves into the background, uniform stellar distribution on a time-scale of $\sim$175 Myr. \citet{bonatto2010} found that the young star clusters from the catalogue of \citet{bica2008} are spatially correlated with one another and with star-forming structures in the ISM. However, the catalogue of \citet{bica2008} was created from earlier catalogues originating from different studies, thus making selection biases and incompleteness difficult to judge. 

Using data from the Visible and Infrared Survey Telescope for Astronomy (VISTA) Magellanic Clouds survey \citep[VMC;][]{cioni2011}, \citet{sun201730dor,sun2017bar,sun2018smc} found that young stellar structures in the Magellanic Clouds have power-law size and mass distributions which can be best explained by a scenario of hierarchical star formation. \citet{zivkov2018} identified 31 pre-main-sequence (PMS) structures in a 1.5 deg$^2$ region just west of the Tarantula Nebula and north of the Bar in the LMC, based on VMC data, with a mass distribution slope which was consistent with hierarchical star formation regulated by turbulence.

\citet{sun201730dor,sun2017bar,sun2018smc} used a contour-based clustering technique to identify the projected two-dimensional (2D) boundaries of groupings of young stars. The technique was first used in nearby galaxies \citep{gouliermis2010,gouliermis2015,gouliermis2017}. This method works by identifying stellar overdensities in a sequence of significance levels above the background stellar density from a surface density map of young stars. It is particularly efficient at partitioning spatial hierarchies. This technique has never been applied to the entire ($\sim$120 square degrees) LMC. 

To better understand the star-formation hierarchy, in this paper we expand the work begun by \citet{sun201730dor,sun2017bar} and exploit the VMC survey's deep near-infrared photometry and high spatial resolution ($<1\arcsec$) to perform a galaxy-wide, contour-based clustering analysis of young stellar structures across the LMC. The VMC photometry in the near-infrared $YJK_\text{s}$ bands is less affected by extinction effects than photometry in optical passbands. Following \citet{sun201730dor,sun2017bar,sun2018smc}, we will refer to all stellar groups, clusters, associations and complexes as young stellar structures. 

In this paper we identify young stellar structures on size scales from a few parsecs to over 1 kpc (bridging from cluster scales to galaxy scales) in the LMC, estimate their parameters and compare with past studies of hierarchical star formation to understand the physical mechanisms behind this seemingly universal architecture. This paper is organised as follows. In Section \ref{sec:two}, we describe the VMC data used, the sample selection of young stars and our identification of statistically grouped young stellar structures. In Section \ref{sec:results}, we show a dendrogram of the second-largest structure identified, we calculate parameters of all young structures, and analyse relevant correlations and distributions. We compare with past studies in the Magellanic Clouds, nearby galaxies and the ISM in Section \ref{sec:discussion}. Section \ref{sec:conclusion} contains our conclusions. 

\section{Data and Young Stellar Structure Identification}
\label{sec:two}

\subsection{VMC Data}
\label{sec:vmc}

\begin{figure}
  \centering
  \includegraphics[width=0.47\textwidth]{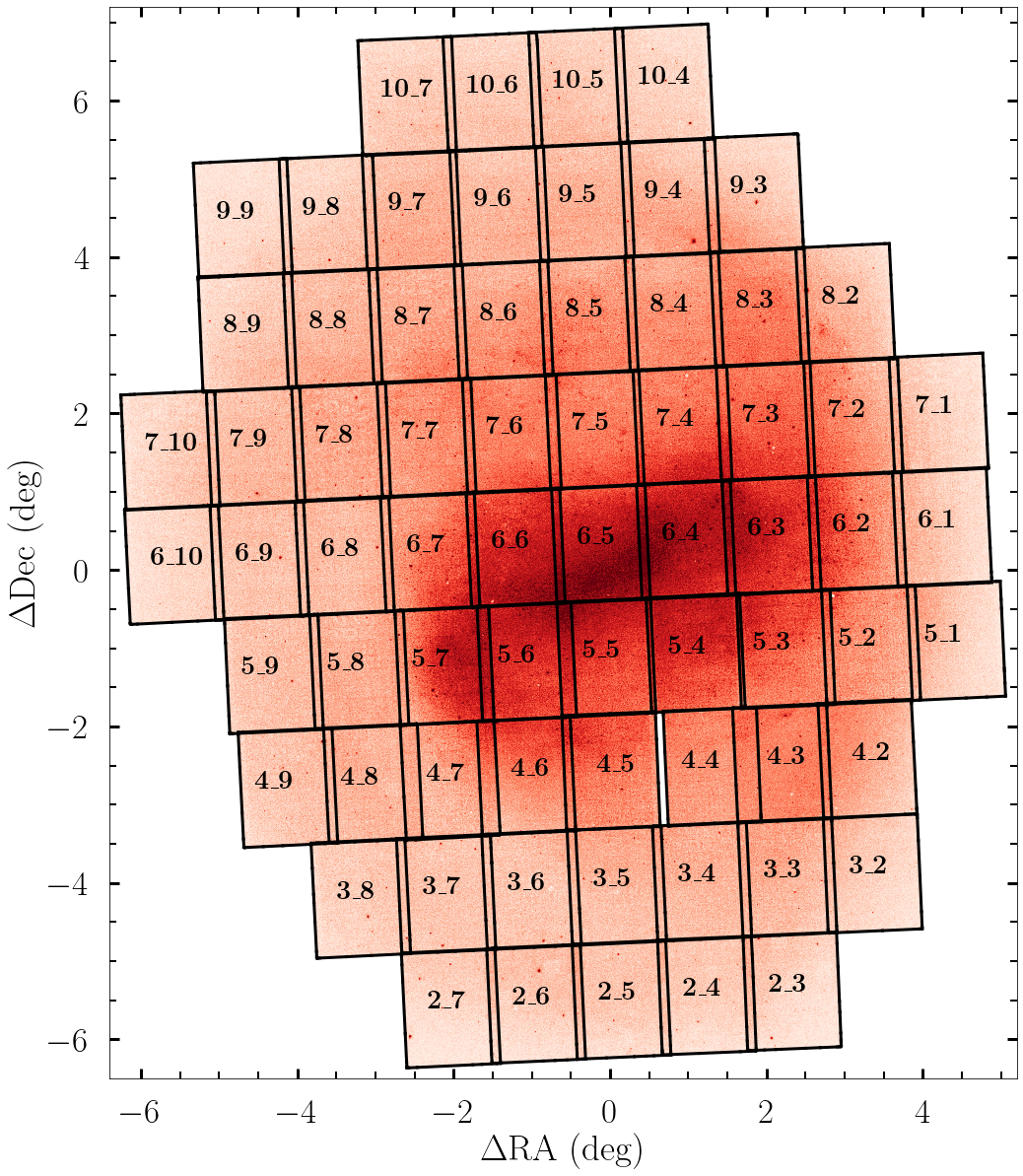}
  \caption{The 68 VMC LMC tiles we study in this paper, overlaid on a stellar density map with sources characterised by $J$ and $K_{\rm s} < 20$ mag. The equatorial coordinates were transformed using a zenithal equidistant projection with the origin at ($80^{\circ}.98, -69^{\circ}.69$), which corresponds to the dynamical centre of young carbon stars in the LMC \citep{wan2020}. }
  \label{fig:tiles}
\end{figure}

The VMC survey \citep{cioni2011} is a near-infrared $YJK_\text{s}$ survey of the Magellanic System covering 170 deg$^2$. The data were gathered between 2009 and 2018 with the 4.1 m VISTA telescope \citep{sutherland2015}, operated by the European Southern Observatory (ESO), using the VISTA infrared camera \citep[VIRCAM;][]{dalton2006}. VIRCAM has 16 detectors arranged in a $4 \times 4$ array; one exposure is a `pawprint'. To cover the gaps between the detectors and ensure contiguous coverage, six pawprints shifted by a specific offset were combined to form a VMC `tile'. Each VMC tile covers $\sim$1.77 deg$^2$. The pawprints are available at both the VISTA Science Archive\footnote{http://horus.roe.ac.uk/vsa/} \citep[VSA;][]{cross2012} and the ESO Science Archive Facility\footnote{http://archive.eso.org/cms.html}. They were reduced using the VISTA Data Flow System \citep{irwin2004,gonzalez2017}, version 1.5, by the Cambridge Astronomy Survey Unit (CASU). Point-spread-function (PSF) photometry was carried out on homogenised, stacked tile images and photometric errors and local completeness have been calculated based on artificial star tests \citep[for more details about the PSF photometry, see][]{rubele2012,rubele2015}. We downloaded the VMC photometric catalogue data from the VSA. Although the VISTA images are publicly available, the PSF photometry used in our work is proprietary to the VMC team, with a data release planned in early 2022. 

There are 68 VMC tiles (shown in Figure \ref{fig:tiles}) covering the LMC, and we include them all in this paper. There are a total of 116,336,429 sources in the 68 tiles. First, we selected stars with $J$ and $K_{\rm s} < 22$ mag and photometric errors $< 0.15$ mag in either filter. After this selection step there were 38,802,981 sources. We next cross-matched adjacent tiles and removed duplicate sources located within 0.5\arcsec of one another, leaving us with 36,030,318 sources. We make additional selection cuts in the next section. Figure \ref{fig:tiles} shows a vertical gap in the stellar distribution. This gap is owing to an observational mistake in which pawprints of tile LMC 4$\_$4 were shifted too far eastwards (in Right Ascension). Complementary VISTA observations are planned with the goal of filling the gap. In this paper, we fill the vertical gap using optical data from the Survey of MAgellanic Stellar History \citep[SMASH;][]{smashdr1,smashdr2}. This process is described in Appendix \ref{appd:gap}. 

\subsubsection{Selection of VMC Upper Main-Sequence Stars} 
\label{sec:ums}

\begin{figure*}
  \includegraphics[width=\linewidth]{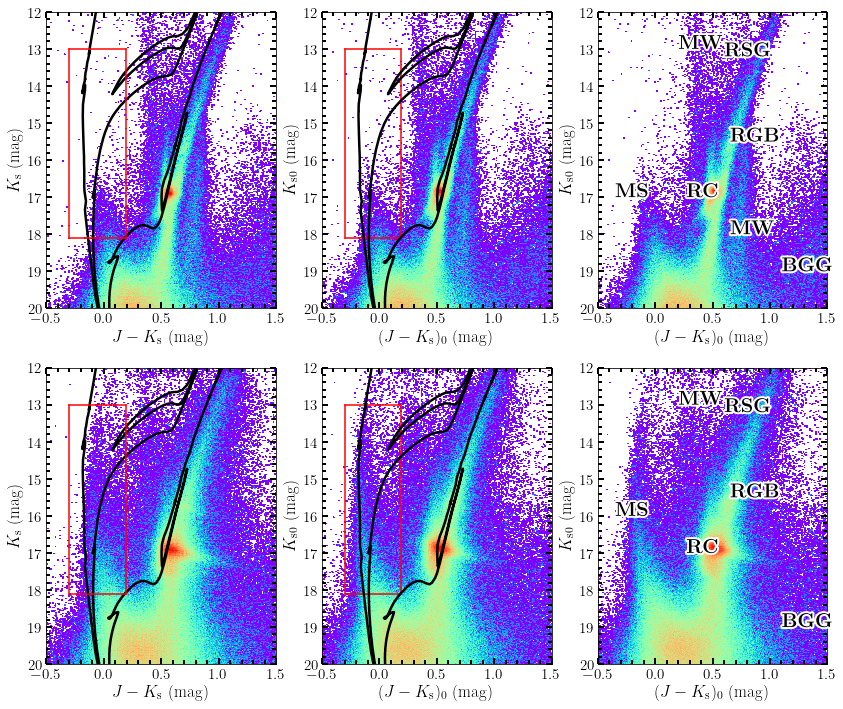}
  \caption{CMDs of VMC tiles LMC 4$\_$6 (top row) and LMC 6$\_$6 (bottom row). All isochrones plotted are of ages $\log(\tau \text{ yr}^{-1}) =$ 7.0, 8.0 and 9.0, in black, from left to right. (Top left) CMD of tile LMC 4$\_$6 with magnitudes uncorrected for extinction, and with isochrones and the UMS selection box overlaid in red. (Top middle) CMD of tile LMC 4$\_$6 with dereddened magnitudes, isochrones and the UMS selection box overlaid in red. (Top right) CMD of tile LMC 4$\_$6 with dereddened magnitudes and components overlaid. (Bottom) As the top row, but for tile LMC 6$\_$6.
  }
  \label{fig:cmds}
\end{figure*}

Figure \ref{fig:cmds} shows $J-K_{\rm s}$ versus $K_{\rm s}$ colour--magnitude diagrams (CMDs) of VMC tiles LMC 4$\_$6 (top panels) and 6$\_$6 (bottom panels). We use this colour combination because it is less affected by reddening than ($Y - K_{\rm s}$). We estimate the prevailing extinction based on the Magellanic Clouds' reddening map provided by \citet{ogle2021}, derived from the properties of red clump (RC) stars and presented in units of $E(V-I)$, with a resolution of $1\arcmin.7 \times 1\arcmin.7$ in the central regions of the LMC. We used $A_{I} = 1.5 E(V-I)$ \citep{ogle2021} and the calibrations of \citet{schlegel1998}'s $E(B-V)$ values -- which we refer to as $E(B-V)_{\text{SFD}}$ -- provided by \citet{schlafly2011}, to express the extinction coefficients in terms of $E(V-I)$. The \citet{schlafly2011} calibration we used is $A_I = 1.505 E(B-V)_\text{SFD}$. Using these two relations and Appendix B of \citet{gonzalez2017}, we derived $A_J = 0.703 E(V-I)$ and $A_{K_{\text{s}}}= 0.307 E(V-I)$. We then dereddened each stellar source by referring to the closest point in the \citet{ogle2021} map.

In Figure \ref{fig:cmds} we also overlaid PARSEC version 1.2S \citep{bressan2012} isochrones of ages $\log(\tau \text{ yr}^{-1}) =$ 7.0, 8.0 and 9.0 and a 
metallicity of [Fe/H] = $-0.3$ dex, which is typical of young OB stars in the LMC \citep{rolleston2002}. In each tile we analyse, the isochrones are shifted by the LMC's distance modulus, $(m-M)_0 = 18.49$ mag \citep{richard2014}. We did not account for the inclination of the LMC's disc and used the same 
distance modulus for all tiles; accounting for the inclination of the disc would slightly shift the position of the isochrones vertically. The isochrones were also shifted from the Vega system to the VISTA system using the relations of \citet{gonzalez2017}: $J_{\text{VISTA}} = J_{\text{Vega}}$ and $K_{\text{s}_{\text{VISTA}}} = K_{{\text{s}}_{\text{Vega}}} - 0.011$. 


The third panel in each row of Figure \ref{fig:cmds} shows the different components visible in the near-infrared CMDs, including Milky Way foreground stars, the main sequence (MS), the RC, the red-giant branch (RGB), red supergiants (RSGs) and background galaxies. To select young MS stars, we looked at the bright young MS, and therefore we selected upper-MS (UMS) stars. We adopted selection criteria based on the PARSEC evolutionary models and similar to those of \citet{sun201730dor,sun2017bar}. We selected UMS stars if they fell into the colour range $-0.3 < J-K_{\rm s} < 0.19$ mag. This was done to differentiate UMS stars from RGB and RC stars. We chose this wider colour range similarly to \citet{sun2017bar}, because although we correct for extinction, the ISM is inhomogeneous across the LMC, which leads to an increase in the width of the MS. We also adopted a magnitude range of $13.0 < K_{\rm s} < 18.1$ mag. The brighter limit at 13 mag allows for some blue-loop stars to fall into our selection criteria, but stars do not spend much time at this evolutionary stage and contamination is therefore minimal. The fainter limit at 18.1 mag was chosen to ensure that all of our UMS stars are younger than 1 Gyr. This is evidenced by the oldest isochrone of 1 Gyr; see Figure \ref{fig:cmds}. After making these cuts, 362,047 sources remain. 

Next, we selected stars with photometric errors less than 0.10 mag, which left us with 355,250 sources. Finally, we made cuts based on the stellar probability parameter, $\tt{Star\_prob}$, and the $J$ and $K_\text{s}$ sharpness indices, $\tt{SHARP}$. As \citet{bell2019} showed (their figure 2), stellar objects should have $\tt{Star\_prob} \geq 0.34$ and $J$ and $K_\text{s}$ sharpness indices $\tt{SHARP} \leq 0.5$. After making these cuts, our selection retained 307,581 UMS stars.

In the region of our selection box shown in Figure \ref{fig:cmds}, contamination owing to foreground Milky Way stars is very small, since they are generally found at colours around $(J-K_\text{s}) \sim 0.7$ mag. We checked the contamination by Galactic sources by cross-matching the VMC LMC sources with {\sl Gaia} early Data Release 3 \citep{gaia2021}. We found in our UMS region that the parallaxes for all matches were $\ll 0.2$ mas, which indicates that contamination by the Milky Way is negligible in our colour--magnitude region \citep[for a more in-depth discussion of foreground contamination in the VMC data, see][]{dalal2021}. Additionally, we checked for contamination using the TRIdimensional modeL of thE GALaxy (TRILEGAL) code \citep{girardi2005}. This code estimates the number of Milky Way stars within a projected sky area. We ran the code on an area of 10 square degrees centred on the LMC. The simulation showed that there is 0.2\% Milky Way contamination in our selection box. Therefore, we conclude that Milky Way contamination is indeed negligible.

\subsection{Stellar Surface Density Map} 
\label{sec:kde}


\begin{figure*}
  \includegraphics[width=1\linewidth]{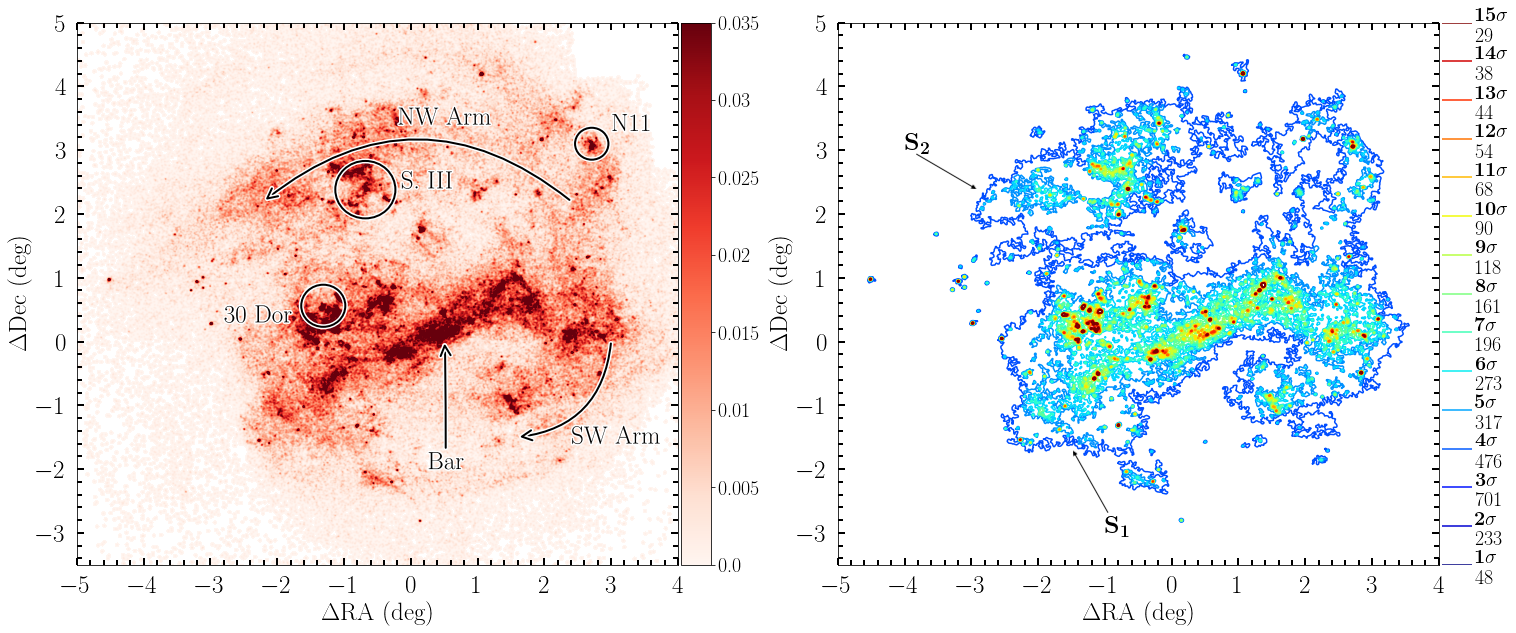}
  \caption{(Left) Surface density map of UMS stars selected from the VMC data. The colour bar is given in units of stars pc$^{-2}$. Labeled are a number of LMC features visible in the young stellar population, as well as three star-forming regions. (Right) Detected young stellar structures coloured by their significance level. The right-hand colour bar is labelled by the significance level and the number of structures contained in each level. The two largest structures are labelled $S_1$ and $S_2$. }
  \label{fig:kde_map}
\end{figure*}

To explore the spatial distribution of UMS stars, we applied kernel density estimation (KDE) -- convolution of the stellar distribution with a Gaussian kernel -- to a 2D binned distribution of the UMS sample. In our application of KDE, the kernel size specifies the resolution of the map. The choice of kernel size is arbitrary and depends on the image resolution, the distance to the galaxy and the size of the objects considered. It is best to experiment to find the optimal kernel size \citep{gouliermis2017,sun2018smc}. The resolution of the VMC survey is 1 arcsec, which corresponds to 0.24 pc at the distance of the LMC. We first grid our UMS into bins of 1 pc $\times$ 1 pc, chosen to be slightly larger than the VMC resolution limit. Next, we experimented with kernel sizes from 5 pc to 20 pc. We found a minimum possible kernel of 10 pc. \citet{sun2017bar,sun2018smc} also found that a kernel of 10 pc represents an optimal balance between resolution and noise. Using these constraints, we created a stellar surface density map. Based on our resolution, we can confidently identify structures larger than 10 pc. The left panel of Figure \ref{fig:kde_map} shows the KDE map with three prominent components of the LMC seen in the young stellar population: the Bar, the Northwestern Arm and the Southwestern Arm. Also pictured are three well-known star-forming regions: 30 Dor, Shapley Constellation III (S.III) and N11; see \citet{dalal2019} for a more in-depth discussion of the morphological features of the LMC's young population. We filled the gap near tile LMC 4$\_$4 with optical SMASH data; see Appendix \ref{appd:gap}. The mean surface density of the KDE map is 0.003 stars pc$^{-2}$, the median surface density is $4\times 10^{-5}$ stars pc$^{-2}$ and the standard deviation is 0.006 stars pc$^{-2}$. These values are similar to those found by \citet{sun2018smc} for the SMC. Most of the values in our map (see the left-hand panel of Figure \ref{fig:kde_map}) and in \citet{sun2018smc}'s map are near zero, but there are some outliers which cause the mean density to be higher than the median.

\subsection{Young Stellar Structure Detection}
\label{sec:detection}



Young stellar structures are identified from the KDE map (see the left-hand panel of Figure \ref{fig:kde_map}) for a range of significance levels above the median density of the KDE map: from 1$\sigma +$ the median stellar KDE density to 15$\sigma +$ the median stellar KDE density, in steps of 1$\sigma$. The median stellar density of the map is effectively zero; therefore, using the median does not affect the statistics of the less populous regions of the LMC, whereas using the mean would. This process resulted in the detection of more than 7000 structures at the 15 significance levels we adopted. To remove spurious detections, we applied two selection criteria: (1) each candidate structure must enclose at least five stars, and (2) every candidate structure on the 1$\sigma$ and 2$\sigma$ levels must contain one or more contours at higher significance levels. The limit of five stars was chosen to coincide with that used by \citet{sun2017bar,sun2018smc}, and it has also been used in similar earlier studies to reduce contamination \citep{bastian2007,bastian2009,gouliermis2017}. After application of these criteria, we identified a total of 2846 structures at the 15 significance levels. The result is shown in the right-hand panel of Figure \ref{fig:kde_map}; the legend on the right shows the different significance levels (1$\sigma$ to 15$\sigma$) and below each level the number of structures identified at each level is indicated. In Appendix \ref{appd:surveys} we compare our catalogue of young stellar structures with similar catalogues from the literature and explain any differences we find. We also estimate that 45\% of our structures were not previously found.

What is the definition of a stellar structure or stellar grouping in the context of our study? As we show in the next sections, our identified structures range in radius from a few parsecs to over 1 kpc. Therefore, our structures include individual stars, young clusters, associations and complexes. The largest identified structures might be associated with galaxy-scale structures or processes. We note that it is difficult to differentiate among clusters, associations, complexes and larger stellar groups. We cannot assess if a group of identified young stars is gravitationally bound, nor whether the stellar group members are located truly close in space. Therefore, if our algorithm identifies a group above the median surface density which passes both of our selection criteria, we refer to it as a young stellar structure or group.

\begin{figure*}
  \includegraphics[width=\linewidth]{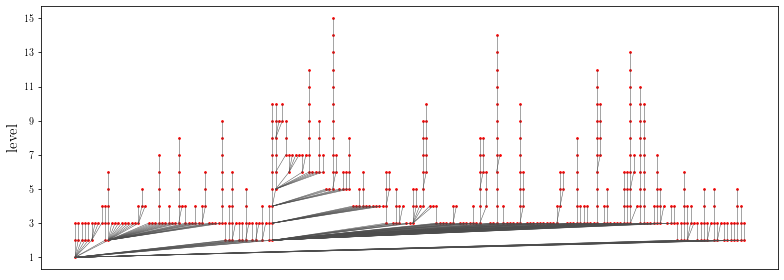}
  \caption{Dendrogram of $S_2$, the second-largest structure we identified, showing the corresponding parent--child relationships. $S_2$ has 451 structures contained within the 14 higher significance levels.  }
  \label{fig:dendro}
\end{figure*}

The right-hand panel of Figure \ref{fig:kde_map} displays the nonuniform design and organisation of the detected young stellar structures. The hierarchical organisation is apparent, especially considering the two largest structures, which we label $S_1$ and $S_2$. To further demonstrate the hierarchical structure, we created a dendrogram of $S_2$, shown in Figure \ref{fig:dendro}. The dendrogram shows the parent--child links between the young stellar structures at the 15 significance levels. 

\section{Results}
\label{sec:results}

\subsection{Parameters of the Young Stellar Structures}
\label{sec:basic}

In the following subsections, we show how we derived structural parameters using the boundaries of the structures shown in the bottom panel of Figure \ref{fig:kde_map}. The basic parameters are the structures' radii $R$ (in pc), perimeters $P$ (in pc) and areas $A$ (in pc$^2$). We account for stellar completeness and calculate the completeness-corrected numbers of stars in each structure, $N_*$. We also calculate the surface density, $\Sigma$, of each structure. Finally, we use the \citet{Larson1981} relation to estimate velocity dispersions ($\sigma_{v}$) and then use it to derive different crossing times ($t_\text{cr}$).

\subsubsection{Fundamental Parameters}
\label{sec:fundamental}
All parameters of the young stellar structures are based on the 2D projected boundaries of the structures identified in Section \ref{sec:detection}. Based on the vertices of each structure, we calculated their areas and perimeters. The radius of each structure is defined as the radius of a circle with the same area as that delimited by each contour.

In order to accurately estimate the number of stars in each structure, we corrected for photometric incompleteness. We followed the same procedure as \citet{sun2017bar}: we first assigned weights to each star, $w = 1.0/\text{min}[f_J,f_{K_{\rm s}}]$, where $f_J$ and $f_{K_{\rm s}}$ are the local completeness levels in the $J$ and $K_{\rm s}$ bands, respectively. The completeness parameters were derived by adding artificial stars to the images, then searching for them in the photometric catalogues \citep[see][]{rubele2012,rubele2015}. Next, we calculated the completeness-corrected number of stars, $N_*$, with
\begin{equation}
 N_* = \sum_i w_i - \Sigma_\text{bg}A,
\end{equation}
where $i$ accounts for each star in the contour, $\Sigma_\text{bg}$ is the median density of the KDE map and $A$ is the area within the contour. The first term corrects for photometric incompleteness and the second term estimates the number of background stars in chance alignment. 


It has been shown by \citet{sun2018smc} that the number of selected UMS stars in young stellar structures is proportional to their total masses to a good approximation (their figure 12). Therefore, we also used $N_*$ as a proxy for mass. Finally, we calculated the stellar surface density, $\Sigma = N_{*}/(\pi R^2)$. 

\subsubsection{Velocity dispersion and crossing time estimated from the Larson relation}
\label{sec:larson}
\citet{Larson1981} identified a power-law scaling relation between velocity dispersion and molecular cloud size:
\begin{equation}
 \sigma_{v} \approx 1.1 (2R)^{0.38}.
 \label{eq:larson_relation}
 \end{equation} 
This relation is known as the Larson relation \citep[for an updated discussion on the exact form of the Larson relation, see][]{cen2021}. It is similar to the Kolmogorov law for incompressible turbulence and is therefore expected to result from turbulence. Studies of young stars in star-forming regions in the Milky Way also show that young groups of stars are generally consistent with the Larson relation \citep{ha2021}. We used the Larson relation to estimate the total three-dimensional (3D) velocity dispersions of the r.m.s. velocity of the internal stellar motions in each structure, $\sigma_v$.

The crossing time is defined as the time it takes a star to travel a distance equivalent to the size of a structure's half-mass size \citep{elmegreen2000time,gouliermis2017,ha2021}:
\begin{equation}
 t_\text{cr} \approx \frac{2R}{\sigma_\text{v}}.
\label{eq:crossing_time}
\end{equation} 
We calculated the crossing times for all identified stellar structures using Equation \ref{eq:crossing_time}. The identified structures are listed in Table \ref{tab:parameters1}. Included are their IDs, significance levels, geometric center coordinates, and physical parameters. Only the first 10 structures are shown, the rest are available online.

\begin{table*}
	\centering
	\caption{The identified young stellar structures and their physical parameters. Columns include ID, significance level ($\sigma$), equatorial coordinates corresponding to the geometric center ($\alpha$(J2000) and $\delta$(J2000) in deg), the radius ($R$ in pc), number of stars ($N_*$), surface density ($\Sigma$ in pc$^{-2}$), velocity dispersion ($\sigma_v$ in km s$^{-1}$), and crossing time ($t_\text{cr}$ in Myr). The first 10 entries are shown here as an example; the complete catalogue is available online.}
	\label{tab:parameters1}
	\begin{tabular}{lrrrrrrrr} 
		\hline 
		ID & Level&$\alpha$(J2000) & $\delta$(J2000) & $R$  & $N_*$ & $\Sigma$& $\sigma_v$ & $t_\text{cr}$  \\
           & ($\sigma$) & (deg) & (deg) & (pc)&    & (pc$^{-2}$)& (km s$^{-1}$)&  (Myr) \\
		\hline
1   & 1  & 93.214 & -68.268 & 40.9.   & 121.1    & 0.023 & 5.6  & 14.6\\
2   & 1  & 90.320 & -67.738 & 26.0.   & 29.0     & 0.014 & 4.7  & 11.0\\
3   & 1  & 90.021 & -68.635 & 24.4.   & 36.6     & 0.019 & 4.6  & 10.5\\
4   & 1  & 89.726 & -68.523 & 41.6.   & 113.0    & 0.021 & 5.6  & 14.7\\
5   & 1  & 89.556 & -68.297 & 22.4.   & 17.1     & 0.011 & 4.5  & 10.0\\
6   & 1  & 89.452 & -68.464 & 34.1.   & 60.6     & 0.017 & 5.2  & 13.0\\
7   & 1  & 89.523 & -68.624 & 26.9.   & 36.9     & 0.016 & 4.8  & 11.2\\
8   & 1  & 89.377 & -69.194 & 36.9.   & 121.0    & 0.028 & 5.4  & 13.7\\
9   & 1  & 84.010 & -67.089 & 1008.6  & 33832.9  & 0.011 & 18.4 & 109.8\\
10  & 1  & 88.954 & -68.205 & 34.9    & 49.9     & 0.013 & 5.3  & 13.2\\
		\hline
	\end{tabular}
\end{table*}

\begin{table*}
	\centering
	\caption{Demographics of the young stellar structures at each significance level. 
	Columns include information about the significance level ($\sigma$), number of structures ($N$), 
	minimum radius ($R_\text{min}$), average radius ($\langle R \rangle$), maximum radius 
	($R_\text{max}$), average number of stars ($\langle N_* \rangle$), average surface 
	density ($\langle \Sigma \rangle$), average magnitude of the fifth brightest star 
	($\langle K_{\text{s0}} \rangle$), average velocity dispersion 
	($\langle\sigma_v\rangle$) and average crossing time 
	($ \langle t_\text{cr} \rangle $) 
	of the young stellar structures at each level.}
	\label{tab:parameters}
	\begin{tabular}{lrrrrrrrrr} 
		\hline
		Level & $N$ & $R_\text{min}$ & $\langle R \rangle$ & $R_\text{max}$ & $\langle N_* \rangle$ & $\langle \Sigma \rangle$ & $\langle\ K_{\text{s0}} \rangle$ & $\langle\sigma_v\rangle$ & $\langle t_\text{cr} \rangle$ \\		($\sigma$) & (structures) & (pc) & (pc) & (pc) & (stars) & (stars pc$^{-2}$) & (mag) & (km s$^{-1}$) &  (Myr) \\
		\hline
		1  & 48  & 19  & 120 & 1839 & 4422$\pm23797$ & 0.01$\pm$0.00 & 16.51 & 6.7$\pm$3.5 & 22.6$\pm$25.9 \\
		2  & 233 & 11  & 40  & 1198 & 667$\pm7509$   & 0.02$\pm$0.01 & 16.93 & 5.1$\pm$1.6  & 13.0$\pm$9.6 \\
		3  & 701 & 3   & 19  & 676  & 164$\pm2115$   & 0.04$\pm$0.02 & 16.76 & 4.0$\pm$1.1  & 8.4 $\pm$5.4 \\
		4  & 476 & 3   & 18  & 488  & 170$\pm1592$   & 0.05$\pm$0.02 & 16.56 & 3.8$\pm$1.2  & 8.0 $\pm$5.4 \\
		5  & 317 & 2   & 20  & 232  & 173$\pm752$    & 0.06$\pm$0.05 & 16.23 & 4.0$\pm$1.2  & 8.5 $\pm$5.2 \\
		6  & 273 & 3   & 18  & 162  & 139$\pm491$    & 0.08$\pm$0.04 & 16.12 & 3.8$\pm$1.1  & 7.9 $\pm$4.5 \\
		7  & 196 & 2   & 18  & 106  & 135$\pm310$    & 0.10$\pm$0.08 & 15.79 & 3.8$\pm$1.1  & 8.0 $\pm$4.2 \\
		8  & 161 & 3   & 16  & 86   & 113$\pm221$    & 0.11$\pm$0.05 & 15.69 & 3.8$\pm$0.9  & 7.7 $\pm$3.4 \\
		9  & 118 & 4   & 15  & 78   & 110$\pm184$    & 0.12$\pm$0.04 & 15.45 & 3.7$\pm$0.9  & 7.5 $\pm$3.1 \\
		10 & 90  & 2   & 15  & 70   & 107$\pm164$    & 0.14$\pm$0.08 & 15.33 & 3.7$\pm$0.9  & 7.3 $\pm$3.0 \\
		11 & 68  & 3   & 14  & 44   & 106$\pm116$    & 0.15$\pm$0.09 & 15.17 & 3.7$\pm$0.8  & 7.2 $\pm$2.6 \\
		12 & 54  & 3   & 14  & 32   & 101$\pm95$     & 0.16$\pm$0.05 & 15.08 & 3.6$\pm$0.7  & 7.0 $\pm$2.4 \\
		13 & 44  & 3   & 13  & 27   & 97$\pm84$      & 0.18$\pm$0.07 & 14.99 & 3.6$\pm$0.7   & 6.8 $\pm$2.2 \\
		14 & 38  & 3   & 11  & 26   & 87$\pm76$      & 0.22$\pm$0.13 & 14.11 & 3.4$\pm$0.7   & 6.4 $\pm$2.0 \\
		15 & 29  & 3   & 11  & 25   & 87$\pm74$      & 0.21$\pm$0.09 & 14.08 & 3.4$\pm$0.6   & 6.4 $\pm$1.7 \\
		\hline    
	\end{tabular}
\end{table*}

\subsubsection{Demographics}
\label{sec:demogrpahics}

Table \ref{tab:parameters} provides an overview of the average parameters at each level. From the lowest (1$\sigma$) to the highest levels (15$\sigma$), the structures' average parameters change: the average $R$, $N_*$ and the average $K_\text{s}$ magnitude of the fifth brightest star decrease, whereas the average $\Sigma$ increases. The velocity dispersions are highest and with the largest standard deviations at the 1,2$\sigma$ levels and lowest and more centred around the mean at the highest significance levels. The crossing times are longest at the 1,2$\sigma$ levels and have smaller standard deviations at the lowest levels. These trends demonstrate the hierarchical nature of the structures, implying that the large lower-density structures contain one or more higher-density groups. 

Estimating the ages of each structure is complicated, and we do not attempt to derive them in this paper. However, we can compare with similar regions in VMC-CMD space defined for the Magellanic Clouds by \citet{dalal2019}. They defined three regions that overlap with our selection box: (1) region A, a population of stars with a median age of $\sim$20 Myr; (2) region B, a population of stars with a median age of $\sim195$ Myr; and (3) region G, a population of stars with a median age of $\sim$81 Myr. Although Figure \ref{fig:cmds} shows that the upper age limit of our sample is 1 Gyr, 90\% of our stars fall within region A. This is evidence that most of the stars we study are around 20 Myr old. In addition, younger stars are spatially more clustered than older stars, and stars older than $\sim$100 Myr exhibit a rather smooth distributions \citep[e.g.,][]{sun2017bar}. Therefore, although our UMS sample covers an age range from a few Myr to just under 1 Gyr, the majority of stars which our detection algorithm selects is younger than 100 Myr. In Section \ref{sec:lmc_env} we discuss age differences across the LMC and how those could affect our analysis.

\subsubsection{Lifetimes}
\label{sec:tcr}

\begin{figure}
  \centering
  \includegraphics[width=0.47\textwidth]{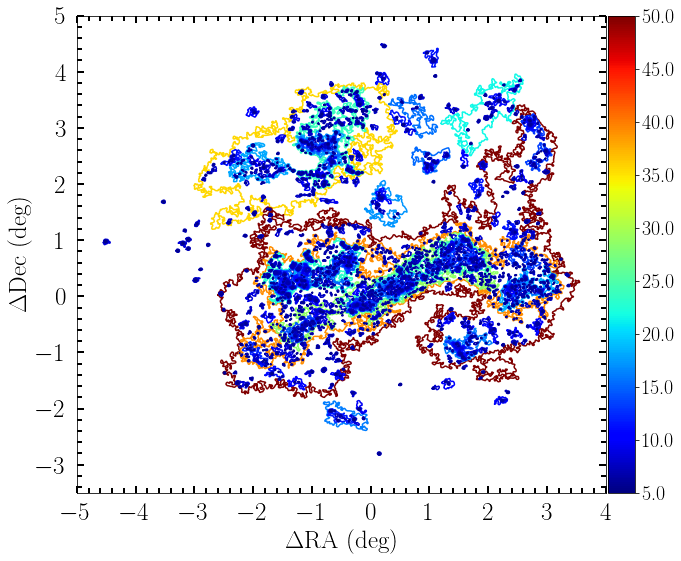}
  \caption{Structure map of the LMC coloured by the corresponding crossing times (in Myr). The colour bar is scaled from 5 Myr to 50 Myr, but the values range from 1 Myr to 160 Myr.}
  \label{fig:tcr}
\end{figure}

Figure \ref{fig:tcr} maps each structure labelled by its derived crossing time. Larger structures have longer crossing times. \citet{allison2009} suggested that all stellar structures will be eliminated into the general background density of a galaxy within a crossing time. Studying the spatial distribution of LMC stars, \citet{bastian2009} supported this suggestion. Therefore, there might be a link between the crossing time of a structure and its lifetime before the stellar structure dissolves into the LMC field. For comparison, the typical crossing time in open clusters is close to 1 Myr \citep{lada&lada2003}. The derived crossing times range from 1 Myr to 160 Myr for the identified young stellar structures. Thus, they will evolve into a smooth distribution within $\sim$100 Myr, consistent with what \citet{bastian2009,sun2017bar} derived.

\subsection{Parameter correlations and distributions}
\label{sec:correl_dist}

\begin{figure*}
  \includegraphics[width=\linewidth]{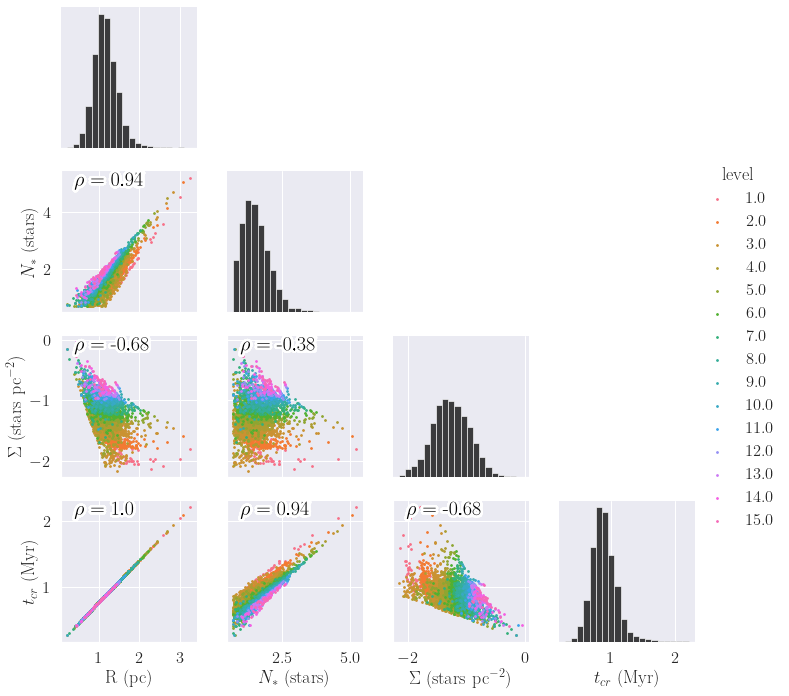}
  \caption{Structure demographics on a logarithmic scale. The diagonals show the histograms of each parameter. Below the diagonals, bivariate scatter plots are shown with the respective Pearson correlation coefficients overplotted in black. The scatter plot data points are labelled by their corresponding significance level; see the legend on the right. }
  \label{fig:pair_plot}
\end{figure*}

In this section we show how the parameters we calculated in Section \ref{sec:basic} are related. Figure \ref{fig:pair_plot} shows pair-wise correlations on a logarithmic scale of the structures' radii, numbers of stars, surface mass densities and crossing times. The diagonal panels show the histograms of each parameter. The lower plots are bivariate scatter plots with the respective Pearson correlation coefficients ($\rho$) overplotted. $R$ and $t_{\rm cr}$ have $\rho = 1$, from the definition of the Larson relation. $R$ is strongly correlated with $N_*$, with $\rho = 0.94$. $N_*$ and $t_{\rm cr}$ have $\rho = 0.94$, as $t_{\rm cr}$ is derived from $R$. The other parameters are not strongly correlated. 

\subsubsection{Perimeter--Area Relation}
\label{sec:perimeter}

\begin{figure}
  \centering
  \includegraphics[width=0.47\textwidth]{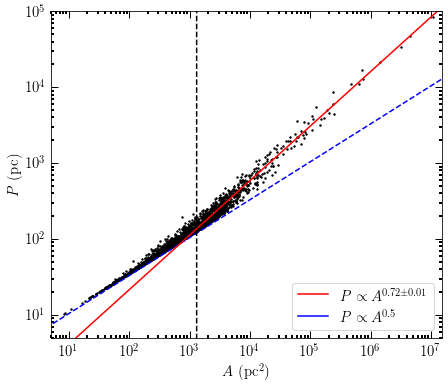}
  \caption{Perimeter--area relation of the young stellar structures. The red solid line is a power-law fit to the structures with area $> 1.3 \times 10^3$ pc$^{2}$. The vertical black dashed line is drawn at the area threshold of $1.3 \times 10^3$ pc$^{-2}$, which corresponds to a structure with a radius of 20 pc. The blue dashed line shows the perimeter--area relationship that is consistent with circular structures.}
  \label{fig:perimeter}
\end{figure}

Figure \ref{fig:perimeter} shows the area ($A$) versus perimeter ($P$) relationship of the young stellar structures. Fitting a power-law slope to the perimeter--area relation assesses whether the shapes of the structures are more regular (circular) or more irregular. In a plane, circles follow the relationship $P \propto A^{0.5}$. For an area $\leq 1.3 \times 10^3$ pc$^{2}$, Figure \ref{fig:perimeter} follows the perimeter--area relationship for circles. We overplot a line with a power-law slope of $\alpha = 0.5$ for points with area $< 1.3 \times 10^3$ pc$^{2}$ to demonstrate that they follow this relationship and also that it cuts off the points above this threshold. The threshold is related to the resolution of the KDE map (10 pc). We choose $A = 1.3 \times 10^3$ pc$^{2}$, corresponding to $R = 20$ pc, to be greater than the resolution of our KDE map, a locus where Figure \ref{fig:perimeter} is not strongly cut off by the perimeter--area relationship pertaining to circles. Beyond this area, we fit a power-law slope to the data using least-squares fitting; we find a best-fitting slope $\alpha = 0.72\pm0.01$. \citet{lovejoy1982} followed Mandelbrot's theory of fractals \citep{mandelbrot1983} to study the geometry of atmospheric clouds on Earth. They used the formula $P \propto A^{D_p/2}$ to parametrise the perimeter--area of the clouds. In this equation, $D_p = 1$ for a circle, increasing towards 2 for more irregular shapes. \citet{falgarone1991} applied this relationship to interstellar molecular clouds. We apply this formula to the slope we fit to Figure \ref{fig:perimeter}. The resulting perimeter--area dimension for the structures is $D_p = 1.44 \pm 0.2$. This indicates that the structures larger than $R$ = 20 pc have very irregular, non-circular shapes.


\subsubsection{Mass--Size Relation}
\label{sec:mass_size}

\begin{figure}
  \centering
  \includegraphics[width=0.47\textwidth]{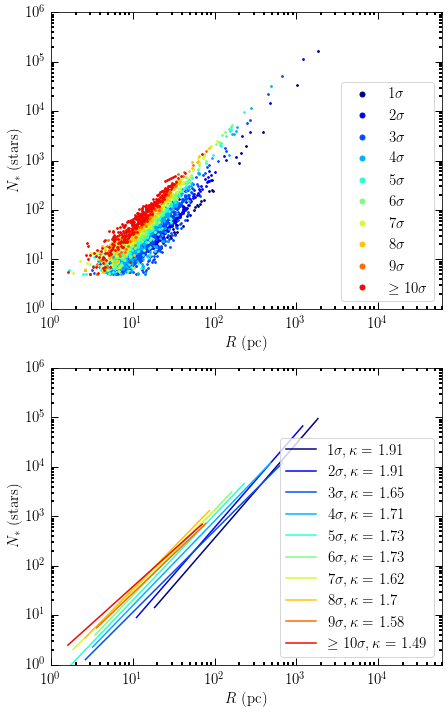}
  \caption{Mass--size relation of the structures. (Top) Scatter plot with different colours labelled by their significance level (shown in the legend). (Bottom) Power-law slopes fitted to the points at each significance level, labelled by their significance levels.}
  \label{fig:mass_size}
\end{figure}

Figure \ref{fig:mass_size} shows the mass ($N_*$) versus size ($R$) relation of the young stellar structures at each significance level. The top panel of Figure  \ref{fig:mass_size} is a scatter plot coloured by each significance level, while the bottom panel shows power-law slopes fitted to the points through least-squares fitting at each significance level. The power-law slopes vary from shallow, around $\kappa = 1.49$, at the lowest significance level to steep, around 1.9, at the highest significance levels. Although the the slope deviates at both ends, it is very similar from 3$\sigma$ to 8$\sigma$. Fitting a line to the entire sample, we find a power-law slope, $\kappa = 1.50\pm0.10$. 

The change in power-law slope at different significance levels can be explained using the 2D fractal dimension, $D_2$. For masses which are hierarchically clustered, the mass--size relation inside a fractal follows the relationship $M \propto R^{D_2}$ \citep{mandelbrot1983}. The 2D fractal dimension is related to, but not the same as, the perimeter--area dimension, $D_p$ (see Section \ref{sec:fractal_dimension}). Applying this to a stellar mass--size relation, we can derive the projected fractal dimension, $D_2 = \kappa$. \citet{feitzinger1987} was the first study to employ Mandelbrot's theory of fractals to stellar populations by determining the fractal dimensions in young star-forming regions in galaxies. They also applied their algorithm to artificial distributions so as to compare their results with those of star-forming regions. They found the uniform artificial distributions to be rhombic and quadratic and have fractal dimensions $\sim$2, whereas star-forming regions have fractal dimensions ranging from $\sim$1.4 to $\sim$1.9. Therefore, stellar structures with a fractal dimension $\sim$2 have a uniform stellar surface density, whereas those with smaller fractal dimensions are characterised by patchy, nonuniform distributions. Applying this framework to the derived slope of our entire sample, we find $D_2 = \kappa = 1.50\pm0.10$. 


\subsubsection{Size, Mass and Surface Density Distributions}
\label{sec:distros}

\begin{figure}
  \centering
  \includegraphics[width=0.475\textwidth]{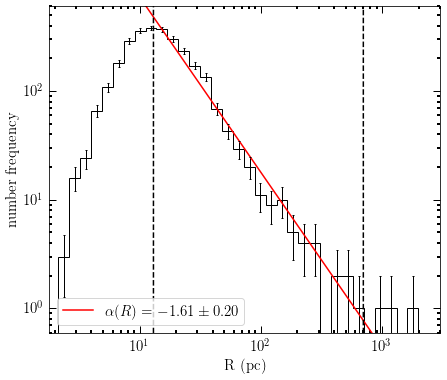}
  \caption{Size distribution of the structures in bins of 0.1 dex. The histogram peaks at 12.5 pc and the mean of the entire sample is $\sim$21.4 pc. The red line is a power-law fit to the data between 13 pc and 700 pc. Error bars represent Poissonian uncertainties. }
  \label{fig:size_dist}
\end{figure}

Figure \ref{fig:size_dist} shows the size distribution of our young stellar structures in bins of 0.1 dex. The distribution peaks at $\sim$13 pc. There is a decline in smaller structures, caused partially by the incompleteness of structures owing to our detection limit of 10 pc and the criterion that each structure must have $N_* > 5$ \citep[see also][]{sun201730dor,sun2017bar,sun2018smc}. Between 13 pc and 700 pc, the size distribution is well-described by a single power law. We hence fit a power law to this size range. We will comment in more detail on this apparent break in the $R$ distribution in Section \ref{sec:compare2mcs}. 

We fitted the power law using least-squares fitting and found $\alpha(R) = -1.61\pm0.20$. We found that changing the bin size of the size distribution does not significantly affect the derived power-law slope. Using the power-law slope of Figure \ref{fig:size_dist}, we can again derive a 2D fractal dimension. For a hierarchically structured sample, substructures follow a cumulative size distribution:
\begin{equation}
N(>R)\propto R^{-D_2},
\label{eq:cumulative}
\end{equation}
where $D_2$ is the fractal dimension \citep{mandelbrot1983,sun2017bar,sun2018smc}. Equation \ref{eq:cumulative} for a cumulative size distribution is equivalent to a differential size distribution of
\begin{equation}
 \frac{{\rm d}N}{{\rm d}\log R} \propto R^{-D_2}.
 \label{eq:differential}
\end{equation} 
Using Equation \ref{eq:differential}, we derive a fractal dimension $D_2 = -\alpha(R) = 1.61\pm0.20$. This slope is consistent with that derived form the mass-size relation, within uncertainties.

As indicated by Figure \ref{fig:size_dist}, there are three structures with $R >$ 700 pc. Considering their boundaries in the context of the LMC, we find that they are similar in size to the overall galactic structure of the LMC. These larger structures are most significantly influenced by galactic processes and therefore do not follow the same power-law size distribution. 

\begin{figure}
  \centering
  \includegraphics[width=0.47\textwidth]{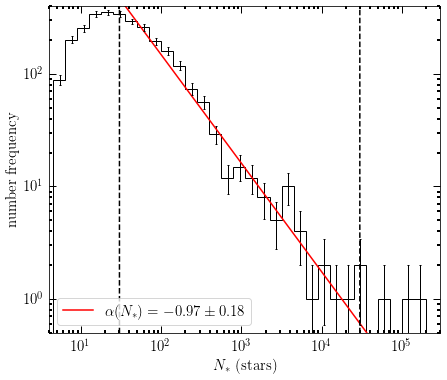}
  \caption{$N_*$ distribution of the structures in bins of 0.15 dex. The red line is a power-law fit to the data covering the range from 30 to 30,000 stars. The peak of the histogram is at 21.45 stars. Error bars represent Poissonian uncertainties.}
  \label{fig:mass_dist}
\end{figure}

Figure \ref{fig:mass_dist} shows the mass ($N_*$) distribution of the young stellar structures. The mass distribution is also incomplete at smaller values. It appears from Figure \ref{fig:mass_size} that once $N_* > 30$ the detection limit of $R = 10$ pc becomes inconsequential. Therefore, we fit a power law from 30 to 30,000 stars. We found that there is no significant change in the fitted power-law slope of the mass distribution when changing the bin size. When plotting the logarithmic mass distribution of masses arranged hierarchically in a fractal-like manner, they should have a power-law slope of $-1$ \citep{elmegreen1996,sun2018smc}. That value is within our margin of error.

There are $\sim$5 structures that have $N_* >$ 30,000 and which are not well-represented by the same power law. This is for the same reason as for the $R$ distribution in Figure \ref{fig:size_dist}: these structures are more likely associated with galactic-scale processes rather than sub-galactic processes. 


\begin{figure}
  \centering
  \includegraphics[width=0.47\textwidth]{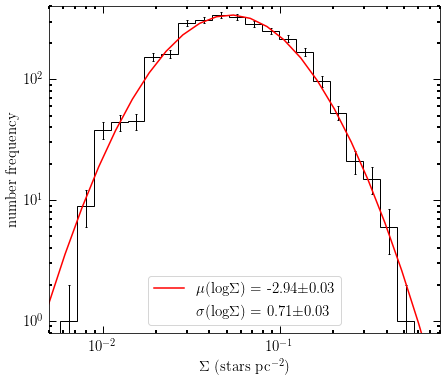}
  \caption{Surface density distribution of the structures in bins of 0.09 dex with a lognormal distribution overplotted in red. The mean and standard deviation of the distribution are included in the legend. Error bars represent Poissonian uncertainties. }
  \label{fig:density_dist}
\end{figure}

Figure \ref{fig:density_dist} shows the distribution of the surface densities ($\Sigma$) for a bin size of 0.09 dex. We fit a lognormal distribution to the data, with the mean and standard deviation listed in the legend. We found that changing the bin size does not affect the distribution. Simulations of supersonic non-gravitating turbulent gas in isothermal environments have a lognormal shape in their column density PDFs \citep{federrath2010,konstandin2012,gouliermis2017}. Therefore, the lognormal shape of Figure \ref{fig:density_dist} indicates that the distribution might be the consequence of turbulence. We elaborate on this further in Section \ref{sec:compare2gas}. 

\subsubsection{On the Fractal Dimension}
\label{sec:fractal_dimension}
Fractal dimensions are used in the literature to show self-similar, irregular structures of groups of stars and gas. There are different methods of obtaining a fractal dimension. $D_p$ and $D_2$ are both types of fractal dimensions, but they are not equivalent. 

As \citet{sun201730dor} noted, $D_2$ is the fractal dimension of our stellar groups' projections onto the 2D plane perpendicular to the line of sight. A suggested relation between the 2D and 3D fractal dimensions is $D_3 = D_2 + 1$ \citep{beech1992}. This relation is used in the literature for gas structures, but it has been shown as improbable for star-forming regions \citep{beech1992,sanchez2005,sanchez2007b,sanchez2008}. Nonetheless, we find that it is useful to derive $D_3$ for purposes of comparison and to differentiate between $D_2$ and $D_p$. Deriving fractal dimensions for gas clouds, \citet{sanchez2005,sanchez2008} showed that $D_3 > D_p + 1$. Our results agree with their findings, since $D_p$ is smaller than both our derived $D_2$ values. Using our two derivations for $D_2$ as well as our derivation of $D_p$, we estimate that $2.34 \leq D_3 < 2.91$. As \citet{sun201730dor} noted, typical values in the ISM are $D_3 = 2.4$. Therefore, the limits of our $D_3$ agree with values found in the ISM more generally.

\subsection{Comparison of young stellar structures in different LMC environments}
\label{sec:lmc_env}

\begin{figure*}
  \includegraphics[width=\linewidth]{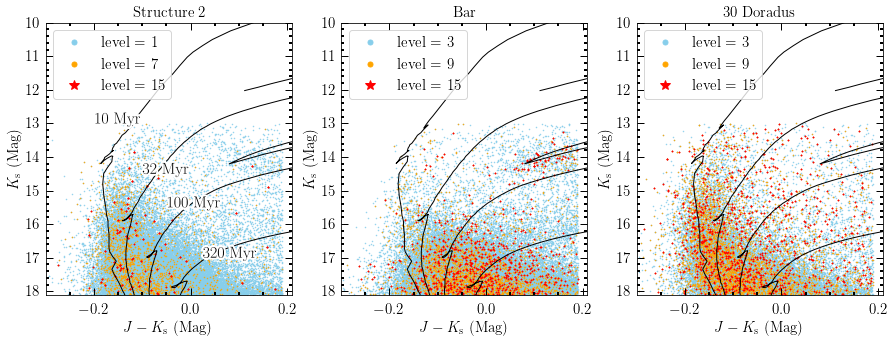}
  \caption{CMDs of stars in three different LMC environments. All plots contain the same isochrones, labelled with their age in the leftmost plot. (Left) CMD of Structure 2, with stars plotted at the 1, 7 and 15 $\sigma$ significance levels. (Middle) CMD of the Bar region, with stars plotted at the 3, 9 and 15 $\sigma$ significance levels. (Right) CMD of the 30 Doradus region, with stars plotted at the 3, 9 and 15 $\sigma$ significance levels.}
  \label{fig:cmds2}
\end{figure*}

In this Section, we inspect three different environments in the LMC: (1) the Bar region; (2) the 30 Doradus region; and (3) Structure 2, also referred to as $S_2$, which contains the Northwestern Arm and Shapely Constellation III; (1) and (2) are located in $S_1$. $S_1$ and $S_2$ are the largest and second largest structures identified, labelled on the right-hand side of Figure \ref{fig:kde_map}.

In Section \ref{sec:demogrpahics}, we commented on the intrinsic difficulties related to deriving ages for our young stellar structures. To give an idea of the age ranges, we plot CMDs of all structures at three different significance levels in these three regions: see Figure \ref{fig:cmds2}. Structure 2 and the 30 Doradus region both show a locus of stars which peak around the isochrones of 10 Myr and 32 Myr, and which show scatter across the entire selection box. However, the Bar region shows the peak shifted redwards towards 100 Myr. There is also shown a cluster of stars around the blue-loop region, corresponding to an age of 100 Myr, in the Bar region. Therefore, the stars in the Bar seem slightly older. Figure \ref{fig:cmds2} displays an expected result based on star-formation studies of the LMC. The Bar region has a larger line-of-sight depth, it is more affected by crowding and contains a tangled mixture of stellar populations. Current and recent star formation are not concentrated in the LMC Bar \citep{harris2009,mazzi2021}. In the LMC, two of the most active star-forming regions are 30 Doradus and the Constellation III region; for example, star-formation activity occurring $<$12 Myr ago is dominated by these two regions \citep{harris2009}. There is some recent star-formation activity in the Bar, but the most active star-forming periods for the LMC Bar occurred instead 5 Gyr, 500 Myr and 100 Myr ago \citep{harris2009}. This can explain the concentration of stars around 100 Myr in the middle panel of Figure \ref{fig:cmds2}. 

\begin{figure}
  \centering
  \includegraphics[width=0.47\textwidth]{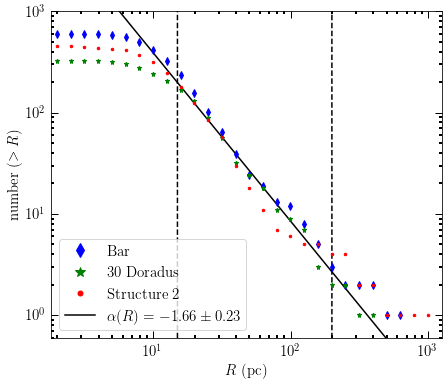}
  \caption{Cumulative size distribution of the Bar, 30 Doradus and Structure 2 ($S_2$) regions in the LMC. A power-law slope is fitted between 15 pc and 200 pc. The power-law slope is the same if we fit all points of the three regions or if we take the average of the power-law slopes of the three regions. The error represents the combined fitting error of the three regions. Error bars are not plotted to avoid confusion.}
  \label{fig:cum_size_dist}
\end{figure}

In order to assess if our results change based on environmental location in the LMC, particularly since there are significant age differences (relative to young stars) across the LMC, we ran our analysis separately for each region. We include the cumulative size distribution of the three regions in Figure \ref{fig:cum_size_dist}. The power-law slope is fitted between 15 pc and 200 pc, and represents the average of three separate power-law slopes (which agree within 1$\sigma$ with each other). The error represents the combined fitting error. We also find that all the power-law slopes and fractal dimensions agree within 1$\sigma$. Therefore, we show that our results do not change based on LMC environment, and we are confident that the difference in stellar populations illustrated in Figure \ref{fig:cmds2} does not affect our conclusions.

\section{Discussion} 
\label{sec:discussion}

\subsection{Comparisons with studies of the ISM}
\label{sec:compare2gas}

Gas structures have size scales from 0.1 pc to 1 kpc and exist in galaxies as a hierarchy of clouds, clumps, cores and filaments \citep{sanchez2008}. Molecular clouds are the birth sites of stars. They show a distinct fractal nature \citep{elmegreen1996,blitz1999,krause2020} with larger clouds hosting smaller, more compact clouds. The largest gas structures are on the same scale as the parent galaxy \citep{elmegreen2000}. They have the same spatial scales as our young stellar structures (a few parsecs to $\geq$1 kpc; see Section \ref{sec:demogrpahics}). 

Gas clouds -- including molecular clouds, high-velocity clouds and H{\sc i} distributions -- have a perimeter--area dimension from 1.2 to 1.5 \citep{sanchez2007b}. Our derived perimeter--area dimension is in this range (see Section \ref{sec:perimeter}). \citet{sanchez2005} found a 3D fractal dimension, $D_3$, of $2.5 \leq D_3 \leq 2.7$ for the Orion A molecular cloud. Considering the H{\sc ii} distributions in galactic discs, \citet{sanchez2008} found 2D fractal dimensions from 1.5 to 2, corresponding to a 3D fractal dimension of 2.73. \citet{sanchez2007b} found a 3D fractal dimension of 2.7 in molecular clouds. These are very similar to the 2D fractal dimensions we derived (see Sections \ref{sec:mass_size} and \ref{sec:distros}) and the 3D fractal dimension ranges we estimated for the young stellar structures in the LMC (see Section \ref{sec:fractal_dimension}). This is evidence the LMC young stellar structures inherited their organisation from the fractal ISM. 

On galactic scales, H{\sc i} is found to have a fractal nature in the LMC \citep{kim2003} and the SMC \citep{Stanimirovic1999}. \citet{elmegreen2001} studied H{\sc i} emission in the LMC and suggested that the hierarchical structure results from supersonic turbulence, based on comparisons with computer simulations \citep{elmegreen1999,pichardo2000}. Turbulence is acknowledged as an important driver of ISM structural evolution. It is believed to play a major role in the creation of molecular clouds and star formation \citep{pingel2018}. Turbulence is created on many spatial scales and cascades to the smallest scales, as shown by the fractal ISM \citep{elmegreen2001}. On larger scales, accretion of the circumgalactic medium and gravitational instabilities probably trigger turbulence \citep{krumholzd&burkhart2016}, whereas on smaller scales stellar feedback in the form of outflows and supernova explosions probably injects energy \citep{padoan2016}. In the SMC, the turbulent properties of H{\sc i} are observed to be homogeneous \citep{nestingenpalm2017}, which seems to be owing to large-scale gravitational driving of turbulence. In contrast, the LMC shows a large diversity in turbulent properties in H{\sc i} \citep{szotkowski2019}. This reveals regions of the LMC that are dominated by gravitational driving of turbulence but also regions on spatial scales smaller than 200--300 pc, where stellar feedback seems to be strongly influencing the turbulent properties.

The fractal nature of these gas structures has been attributed to the physical processes that create them in the ISM. Observations of interstellar gas and dust show a fractal nature in their distributions, which can be compared with theory to indicate that turbulence is an important process in the ISM. Simulations of compressively driven supersonic turbulence imply a 3D fractal dimension of $D_3 = 2.7$ \citep{federrath2009}. This is in the range of our derived $D_3$ values (see Section \ref{sec:fractal_dimension}) and indicates that the clustering of star formation in the LMC on scales from 10 pc to $\sim$700 pc is driven by supersonic turbulence.

It is important to note that hierarchical structure arises from both top-down (fragmentation of molecular clouds) and bottom-up (injection of energy by stellar winds and supernovae) processes. Therefore, the smallest structures we have observed could follow more closely the fractal structure of self-gravitating molecular clouds, whereas the larger structures could possibly follow more closely the larger-scale ISM structure dominated by the top end of the supersonic turbulence spectrum. 

Studies of molecular clouds show that those that follow a lognormal surface density distribution are dominated by supersonic turbulence \citep{padoan2002,elmegreen2011b,gouliermis2017}. In contrast, molecular clouds that have a surface density distribution with a power-law tail are more strongly controlled by self-gravity. We showed that the surface density distribution of the young stellar structures follows a lognormal distribution (Section \ref{sec:correl_dist} and Figure \ref{fig:density_dist}). However, there is the caveat that the absence of a power-law tail in our size distribution could be attributed to the constraints of our study, namely the resolution of our KDE map of 10 pc. We might not be able to detect the effects of self-gravity in our study, because we cannot detect the smallest, densest structures. The lognormal shape of our surface density distribution is a further demonstration that the congregation of star formation in the LMC is regulated by supersonic turbulence.

\subsection{Comparisons with past work on the Magellanic Clouds}
\label{sec:compare2mcs}

\citet{gouliermis2003} expanded on the stellar detection method introduced by \citet{gouliermis2000} and applied it to a $6.5^{\circ} \times 6.5^{\circ}$ area centred on the LMC Bar. With a detection limit of 20 pc, they detected 494 systems. They separated the systems into bound, intermediate and unbound using their calculated central densities. Their unbound systems are described as representing stellar associations. They found a mean diameter of these of 85 pc. The mean size or peak in the size distribution of a young stellar system is important, because it could represent a universal scale for star formation. We will discuss the presence or absence of this characteristic scale in Section \ref{sec:compare2others}. 

\citet{bastian2009} studied the spatial distributions of different stellar populations in the LMC using the optical Magellanic Cloud Photometric Survey \citep{MCPS1997}. They used a fractured minimum spanning tree approach to break their sample of stars into separate groups. For this, a breaking distance was applied in which all edges of the tree longer than the breaking distance were removed. They applied 20 different breaking distances. Their detection limit was 1.5 pc. From this approach they created a catalogue of objects compiled from every breaking distance applied and removed duplicates. Using Two Point Correlation Functions (TPCF) and the $Q$ parameter proposed by \citet{cartwright2004}, which can differentiate between centrally concentrated and fractal distributions, they derived a 2D fractal dimension (see Table \ref{tab:compare_mcs}). From their results, they hypothesised that the stellar structure evolution in the LMC may be most affected by general galactic dynamics: the galactic velocity dispersion may erase substructures and render the stellar distribution uniform. They conjectured that all stellar structures would be eliminated in a crossing time. From our results, we suggest that the young stellar structures are not most affected by general galactic dynamics. It is not the motion across the LMC that erases the young structures. Instead, it is their internal disequilibrium that causes them to disperse and then local dynamical effects that smooth the stellar distribution. Additionally, the LMC's global velocity dispersion is dominated by older stars which acquired deviant velocities as a result of dynamical relaxation. Our young stellar structures do not experience global stirring; they form part of an orderly, rotating disc \citep{gaia2021}. 

Plotting the cumulative size distribution (their figure 1) using four different breaking distances, \citet{bastian2009} found no evidence of a characteristic scale of star formation. They found that the breaking distance used corresponds to the mean of the distribution in a one-to-one manner. Therefore, they concluded that there is no evidence of a characteristic scale of star formation in the LMC and that, in general, characteristic size scales found in different studies may be the result of the selection of a single breaking distance or size scale used. 

For the cumulative size distribution for their total sample, they note the existence of an excess at radii $> 400$ pc. This break is similar to that seen in the autocorrelation function of young stars in the LMC \citep[1 kpc;][]{odekon2008} and the scale where a break in the H{\sc i} distribution is observed: 100--290 pc \citep{elm&elm2001,padoan2001,kim&park2007}. It has been suggested that these observed breaks are owing to the disc's geometry, for example the scale height of the disc, which is observed to have a varying thickness from $\sim$50 pc to $\sim$ 400 pc \citep{szotkowski2019}. We find a similar `break' in our size distribution (Figure \ref{fig:size_dist}) around 300--700 pc. We suggest that this is related to the scale height of the LMC, the size of the galaxy, and the disc structure. Owing to the break of $\sim$700 in our size distribution, we infer that around this scale a transition occurs: below this scale supersonic turbulence dominates, whereas above this scale global galactic structure dominates.

\citet{sun201730dor,sun2017bar} studied young stellar structures in two VMC tiles which contain 30 Dor and the Bar star-forming complex. They found structures ranging from a few parsecs to 100 pc in size. \citet{sun2018smc} studied young stellar structures in an area of $\sim28$ square degrees centred on the SMC using 16 VMC tiles. They identified 556 structures on 15 significance levels, whereas we identified 2846 structures on 15 significance levels. Their structures range in size from a few parsecs to 1000 pc. We compare our results with \citet{sun201730dor,sun2017bar,sun2018smc} in Table \ref{tab:compare_mcs}. We find similar results for the LMC and SMC. In their size distribution, \citet{sun2018smc} found that the peak was at $\sim$10 pc, similar to the peak we found at $\sim$12 pc. One important difference is the fitting range they used for their size distribution: from 10 pc to 100 pc. Their upper bound is lower owing to the smaller size of the SMC. A comparison of our results shows that young star-forming regions with ages from a few Myr to 100 Myr are similar in the LMC and the SMC in terms of their properties (size, mass, area, etc.) despite differences in their physical and chemical environments. 

\citet{zivkov2018} studied 31 PMS structures in the LMC using VMC data. Fitting a power law to the number distribution of their structures, they derived a slope of $-0.86\pm0.12$. This agrees with our value within the errors. 

Table \ref{tab:compare_mcs} compares the derived parameters $D_2$, $D_p$, the power-law fit to $N_*$ and the shape of the surface density distribution from this work and four other similar studies of the Magellanic Clouds. In general, the results are very similar. We suggest that the differences in $D_2$ may be owing to the method of determination (power-law fit versus the TPCF/$Q$ parameter). When fitting a power law, the results depend on whether the cumulative or differential size distribution is considered. Moreover, the slope depends on the size of the binning scheme and the mathematical fitting algorithm. The derived exponents for $D_2$ are all $<2$, which is expected for a fractal distribution of stars. 

\begin{table*}
	\centering
	\caption{Fractal dimensions and power-law slopes for young stars in the Magellanic Clouds. Columns include data about the region of interest and reference and derived $D_2$, $D_p$, a power-law fit to the number of stars ($\alpha (N_*)$) and the shape of the surface density distribution ($\Sigma$ Dist.). $D_2$ values are further labelled by a superscript which indicates the derivation method: $^1$mass--size relation; $^2$size distribution; $^3$TPCF/$Q$ parameter.}
	\label{tab:compare_mcs}
	\begin{tabular}{lrrrr} 
		\hline
		Galaxy/Region & $D_2$ & $D_p$ & $\alpha (N_*)$ & $\Sigma$ Dist. \\
		\hline
		LMC: 30 Doradus \citep{sun201730dor} & $1.6\pm0.3^2$ & $--$ & $--$ & $--$ \\
		LMC: Bar \citep{sun2017bar} & $1.5\pm0.1^1$; $1.5\pm0.1^2$; $1.75\pm0.01^3$  & $--$ & $-1\pm0.1$ & lognormal \\
		LMC: This work & $1.50\pm0.10^1$;$1.61\pm0.20^2$  & $1.44\pm0.20$ & $-0.97\pm0.18$ & lognormal \\
        SMC: \citep{sun2018smc} & $1.48\pm0.03^1$;$1.4\pm0.1^2$  & $1.44\pm0.20$ & $-1\pm0.1$ & lognormal \\
        LMC: \citep{bastian2009} & $\sim1.8^3$  & $--$ & $--$ & $--$ \\
		\hline    
	\end{tabular}
\end{table*}

\subsection{Comparisons with past work on other galaxies}
\label{sec:compare2others}

\begin{table*}
	\centering
	\caption{Stellar structures detected in other galaxies. Listed are the galaxy and, in parentheses, whether or not the study encompasses the entire galaxy, the distance, the number of stellar structures detected, the resolution of the study (if known), the $R$ distribution peak (converted to radius if indicated originally in units of diameter), $D_2$ and (if relevant) the fitting range, $D_p$ and the shape of the surface density ($\Sigma$) distribution. The superscripts associated with the $D_2$ values indicate the derivation method: $^1$mass--size relation; $^2$size distribution; $^3$autocorrelation function. The subscripts adjacent to galaxy names indicate the reference: $_1$this work; $_2$\citet{sun2018smc}; $_3$\citet{gouliermis2015}; $_4$\citet{gouliermis2017}; $_5$\citet{rodriguez2016}; $_6$\citet{rodriguez2018}; $_7$\citet{rodriguez2019}; $_8$\citet{rodriguez2020}. }
	\label{tab:galaxy_wide}
	\begin{tabular}{lrrrrrrrr} 
		\hline
		Galaxy(-Wide?)  & Dist. &  Structures & Res. (pc) & $R$ Peak (pc) & $D_2$ & $D_p$ & $\Sigma$ Shape\\
		\hline
		LMC$_1$ (yes)& 50 kpc &  2846 & 10 & 13 & $1.50^1$;$1.61^2$ (13--700 pc)  & $1.44$ & lognormal\\
		SMC$_2$ (yes)& 62 kpc &  556 & 10 & 10 & $1.48^1$;$1.40^2$ (10--100 pc)  & $1.44$ & lognormal\\
		NGC 6503$_3$ (yes)& 5 Mpc & 244 & 80 & 65 & $1.6^1$;$1.5^2$ (65--1000 pc);$1.7^3$  & $--$ & $--$\\
		NGC 1566$_4$ (yes)& 10 Mpc & 890 & 67 & 63 & $1.54^1$;$1.8^2$ (63--1000 pc) & $--$ & lognormal, bimodal\\
		NGC 300$_5$  (no) & 1.93 Mpc & 1147 & $--$ & 25 & $--$ & $1.58$ & $--$\\
		NGC 253$_6$  (no) & 3.56 Mpc & 875 & $--$  & 40 & $--$ & $1.50$ & $--$\\
        NGC 247$_7$  (no) & 3.60 Mpc & 339 & $--$  & 55 & $\sim1.8^2$ (30--150 pc) & $1.58$ & $--$\\
        NGC 2403$_8$ (no) & 3.18 Mpc & 573 & $--$  & 40 & $--$ & $1.64$ & $--$\\
		\hline    
	\end{tabular}
\end{table*}

The parameters we will pay most attention to in this section are the fractal dimensions ($D_2$ and $D_p$) and the mean/peak (which are closely related) of the size distributions. As stated in Section \ref{sec:compare2mcs}, the mean/peak of the size distribution in studies of young stellar structures or stellar associations can be indicative of a universal, characteristic scale of star formation. 

First, as regards the characteristic scale (mean/peak of the size distribution), \citet{melnik&efremove1995} studied OB associations within 3 kpc of the Sun and found a maximum in the size distribution at 15 pc and an average radius of 20 pc. \citet{bresolin1998} studied the OB associations of seven spiral galaxies and found that the size distribution peaked at radii of 20--40 pc. \citet{pietrzynski2005} compared OB association parameters of 22 galaxies and found a peak of the size distribution at 25--55 pc with an average of 20--90 pc. \citet{efremov&ivanov&nikolov1987} and \citet{ivanov1996} found that young stellar associations have average radii of $\sim$40 pc. Owing to the abundance of stellar association studies, a radius of $\sim$40 pc has been discussed as a characteristic galactic scale for star formation \citep[see][]{gouliermis2011}. However, it is important to note that these studies only considered unbound stellar groups as stellar associations, not smaller, more compact systems in the star-formation hierarchy. 
 
\citet{gouliermis2015} and \citet{gouliermis2017} expanded on the young stellar structure method used by \citet{gouliermis2003}. The parameters of their studies are listed in Table \ref{tab:galaxy_wide}. They found peaks in their size distributions of 63--65 pc. This is larger than those found in this paper and by \citet{sun2018smc}. We suggest that this is because of the distances to the galaxies considered and the resolution limit of their studies (80 and 67 pc, respectively). Their derived $D_2$ values are very similar to ours and are based on a similar fitting range. \citet{gouliermis2017} found a bimodal surface density distribution. They noted that the first node is composed entirely of structures found at the first significance level. This is different from our unimodal surface density distribution (see Figure \ref{fig:density_dist}), but it did not lead them to reach different conclusions from ours. 

\citet{gouliermis2017} also used the viral theorem to estimate the velocity dispersions and crossing times of their structures. Because many of their structures are unbound, their crossing times are upper limits to the true values, whereas the derived velocity dispersions are lower limits to the true values. From their 1$\sigma$ to 12$\sigma$ levels, they found an average range of 0.5--0.8 km s$^{-1}$ for the velocity dispersion and an average range of 71--406 Myr for the crossing times. The average values we estimated for the velocity dispersion (see Table \ref{tab:parameters}) are larger by an order of magnitude and display a different trend: the lowest significance-level structures (on the 1 and 2$\sigma$ levels) have the highest velocity dispersions. We assert that using the Larson relation leads to more accurate velocity dispersion estimates for young stellar structures. One reason is that studies show that velocity dispersions for younger clusters increase as they age \citep{ramirez2021}. Another is that our values are similar to the velocity dispersion for RSG stars in the young LMC cluster NGC 2100 \citep{patrick2016}. 

\citet{rodriguez2016,rodriguez2018,rodriguez2019,rodriguez2020} studied four galaxies in the Local Group. The parameters of their studies are listed in Table \ref{tab:galaxy_wide}. It is important to note that these four studies are not galaxy-wide studies; they instead looked at one or more images covering part of the galaxy. They used a path-linkage criterion technique to detect young stellar groups and create a size distribution. They also used KDE maps. They found similar $D_2$ and $D_p$ values to ours and the other studies included in Table \ref{tab:galaxy_wide}.

\citet{grasha2017} studied six local (3--15 Mpc) star-forming galaxies. With a completeness limit of 1 pc, they studied 3685 young stellar groups. From application of the TPCF, they found an average $D_2$ of 1.17, but they emphasised that their results vary for each galaxy and that the different clustering strengths imply different fractal dimensions and different star-formation hierarchies for each galaxy. They found a characteristic radius from 50 pc to 150 pc. Their average $D_2$ is much smaller than those listed in Tables \ref{tab:compare_mcs} and \ref{tab:galaxy_wide}. This could be owing to the different method they used to estimate $D_2$. The characteristic size is larger than the mean and peak of our size distribution, but similar to many of those listed in Table \ref{tab:galaxy_wide}.

\citet{mondal2019} studied the dwarf irregular galaxy IC 2574 (3.79 Mpc) and identified 419 bright far-UV groups. With a resolution limit of 15 pc, they found that their groups have radii of 15--285 pc and they identified some of them with stellar associations. \citet{mondal2021} completed a similar study for the flocculent spiral galaxy NGC 7793 and identified 2046 young star-forming clumps. With a detection limit of 6.8 pc, they found radii between 12 pc and 70 pc. These two studies have a similar lower bound to their young star-forming regions, but a lower upper bound than ours. This is owing to the sizes of the galaxies considered and the fact that they use far-UV data; in addition, their definition of a star-forming group is different from our young stellar structures. 

\citet{gusev2014} studied NGC 628 (7.2 Mpc) and, using a detection method similar to \citet{gouliermis2010}, detected 297 young stellar structures with a detection limit of 50 pc. They found a bimodal size distribution, with the highest peak at a radius of $\sim$30 pc and a lower peak at 300 pc. Considering the cumulative size distribution, they fitted a power law. They found a power-law slope of $-1.5$ in three radial ranges: (1) 23--43 pc; (2) 95--300 pc; and (3) 325--450 pc. They stated that they found three characteristic sizes of young star groups: (1) OB associations with mean radii of 33 pc; (2) stellar aggregates with mean radii of 120 pc; and (3) stellar complexes with mean radii of 292 pc. Therefore, they did not find a continuous power-law slope in the range of 50--1000 pc. In contrast, we found a continuous power law covering scales of 13--700 pc and a single peak in our size distribution. These differences could be owing to the detection method, the different types of distribution used to estimate the power-law slopes or the fitting algorithm used. No comparable study found similar breaks in their cumulative or differential size distributions. 

Comparing results from the LMC, SMC and the other galaxies referred to in this subsection, we can see if the hierarchical properties depend on metallicity and/or environment (e.g., starburst, galaxy interactions, etc.). It seems that there is no obvious dependence. Therefore, regardless of which physical processes take place on large galactic scales or which microphysics controls the small scales, the clustering of star formation between the scale of clusters and galaxies seems to be universally controlled by turbulence-driven hierarchical star formation.


\section{Conclusion} 
\label{sec:conclusion}
We present the highest-resolution study to date of the observed hierarchical architecture of groups of young stars in the LMC. This is the first such study done covering an area of $\sim$120 square degrees across the LMC. The main results of the study are as follows:
\begin{enumerate}
  \item We identified 2846 young stellar structures on 15 significance levels from a KDE map (with a resolution of 10 pc) from young stars selected from 68 VMC tiles. 
  \item We showed the hierarchical organisation of the young stellar structures by displaying a dendrogram of the second-largest structure.
  \item We calculated parameters of the young stellar structures: radius, area, perimeter, surface density, number of stars, crossing time and velocity dispersion.
  \item The structures range in size from a few parsecs to 1000 pc, with a mean radius of $\sim$21 pc.
  \item The velocity dispersions range from 1.7 to 23 km s$^{-1}$.
  \item The crossing times range from 2 to 160 Myr. 
  \item From the perimeter--area relation, we derived a perimeter--area dimension of $D_p = 1.44\pm0.20$.
  \item From the mass--size relation, we derived a 2D fractal dimension of $D_2 = 1.50\pm0.10$.
  \item The size distribution of the structures has a peak at $\sim$13 pc. From the size distribution, we derived a 2D fractal dimension of $D_2 = 1.61\pm20$ in the range of 13--700 pc.
  \item From the $N_*$ we fitted a power law from 30 to 30,000 stars, with a slope $\alpha(N_*) = -0.97\pm0.18$. 
  \item We showed that the surface density distribution fits that of a lognormal distribution. 
\end{enumerate}

Our distributions and fractal dimensions ($D_2$ and $D_p$) are similar to those found previously in the LMC and other galaxies. There are strong similarities in the fractal dimensions and the surface density distribution to those of structures in the ISM and numerical simulations. Therefore, we conclude that our identified stellar structures inherited their fractal nature from the ISM and that supersonic turbulence is a dominant mechanism in generating this architecture. Our results support the explanation of turbulence-driven hierarchical star formation across different galactic scales. Our findings suggest that star formation in the LMC, on scales from 10 pc to $\sim$700 pc, is scale-free and forms a continuum of star-formation structures. 

\section*{Acknowledgements}

This project has received funding from the European Research Council (ERC) under the European Union's Horizon 2020 research and innovation programme (grant agreement no. 682115). This research was also supported in part by the Australian Research Council Centre of Excellence for All Sky Astrophysics in 3 Dimensions (ASTRO 3D), through project number CE170100013. We would like to thank the Cambridge Astronomy Survey Unit (CASU) and the Wide Field Astronomy Unit (WFAU) in Edinburgh for providing the necessary data products under the support of the Science and Technology Facilities Council (STFC) in the U.K. This study is based on observations made with VISTA at the ESO/La Silla Paranal Observatory under programme ID 179.B-2003. Finally, this project has made extensive use of the \texttt{Python} packages: Astropy \citep{Astropy18}, matplotlib \citep{hunter2007}, NumPy \citep{numpy}, pandas \citep{pandas} and SciPy \citep{scipy}. 

\section*{Data Availability}

The entire young stellar structure catalogue is available at \url{https://github.com/amyelizmiller/VMC_LMC_hierarchical_star_formation}. The VMC images are publicly available, but the PSF VMC photometry used in this work are propriety to the VMC team, with a data release planned in early 2022; see \url{https://www.eso.org/sci/publications/announcements/sciann17232.html}. The SMASH data used are part of DR2 and are available at \url{https://datalab.noao.edu/smash/smash.php#SecondDataRelease}.
\bibliographystyle{mnras}
\bibliography{references}

\appendix

\section{Filling in the Gap with SMASH data}
\label{appd:gap}

\begin{figure}
  \centering
  \includegraphics[width=0.47\textwidth]{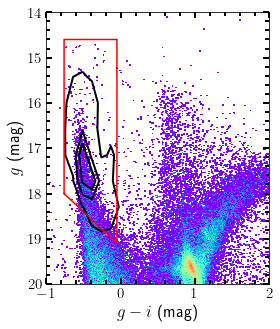}
  \caption{SMASH CMD in the rectangular region around the gap of VMC data. The black contours show the density of the cross-matches of UMS stars. The red polygon is the region we used to select stars.}
  \label{fig:smash}
\end{figure}

\begin{figure}
  \centering
  \includegraphics[width=0.47\textwidth]{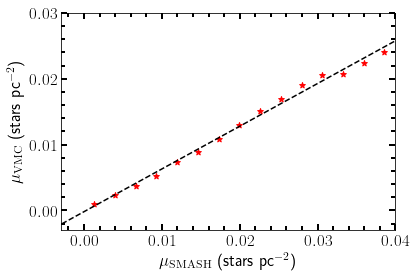}
  \caption{Pixel-to-pixel comparison of the pixels of the SMASH and VMC KDE maps. The points represent mean $\mu_\text{VMC}$ values in 15 equally spaced $\mu_\text{SMASH}$ bins.}
  \label{fig:pixel}
\end{figure}

Figure \ref{fig:tiles} (in Section \ref{sec:vmc}) shows an absence of data centred at ($\Delta$RA, $\Delta$Dec) = $(0.65,-2.48)$ with a width of 0.108$^{\circ}$ and a height of 1.445$^{\circ}$. To fill this gap, we followed the method of \citet{sun2018smc}. They filled a gap in VMC data of the SMC using $UBV$ photometric data. We filled the gap in the LMC coverage using optical data from SMASH Data Release 2 \citet{smashdr2}. SMASH data were gathered with the Dark Energy Camera \citep[DECam;][]{flaugher2015} and cover the Magellanic System in fields of $\sim$3 deg$^2$ in the $ugriz$ passbands. A complete description of the survey strategy and data reduction can be found in \citet{smashdr1,smashdr2}. To use SMASH data, we first defined a rectangular region around the gap with dimensions of 2$^{\circ} \times 1.26^\circ$. This region encloses a region of 0.1 pc to 0.2 pc around the gap; in this region there are 1119 UMS stars selected from the VMC data around the gap. We identified this region of sky in the SMASH survey. To select point-like sources from the SMASH data, we made cuts using three morphology parameters: $\tt{CHI} < 3$, $|\tt{SHARP}| < 1$ and $\tt{PROB} > 0.8$. Using this selection, we cross-matched with UMS stars in the same rectangular region. There were 1094 matches in this region. Figure \ref{fig:smash} shows a CMD of SMASH data in our rectangular region, with the black contours representing the density of cross-matches and the red polygon showing our selection of UMS stars. There are 1937 SMASH stars in this region of CMD space.

\begin{figure}
  \centering
  \includegraphics[width=0.47\textwidth]{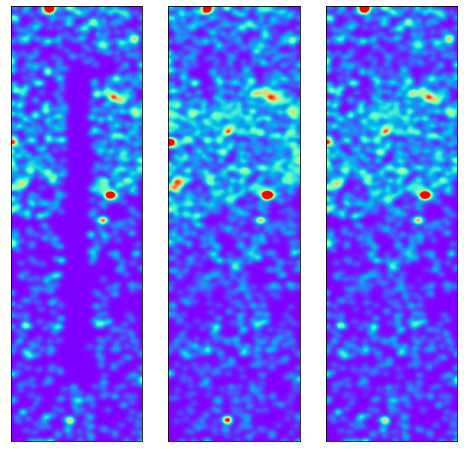}
  \caption{KDE maps of the rectangular region centred at ($\Delta$RA, $\Delta$Dec) = $(0.65,-2.48)$ with size 2$^{\circ} \times 1.26^\circ$. (Left) KDE map of the VMC data. (Middle) KDE map of the calibrated SMASH data. (Right) Combined map of the VMC and SMASH data.}
  \label{fig:combo}
\end{figure}

We constructed surface density maps of the rectangular region using both VMC and SMASH data. The left-hand panel of Figure \ref{fig:combo} shows the VMC KDE map of this region. We calibrated the pixels of the SMASH KDE map, $\mu_\text{SMASH}$, by doing a pixel-to-pixel comparison, shown in Figure \ref{fig:pixel}. We fitted a line to this comparison and found $\mu_\text{VMC} = 0.65 \mu_\text{SMASH}$. The middle panel of Figure \ref{fig:combo} shows the calibrated SMASH KDE map. To combine these two maps, we used equation (2) of \citet{sun2018smc}: $\mu_\text{comb} = \mu_\text{VMC} \times w + \mu_\text{SMASH} \times (1 - w)$, where $w$ is a weight assigned to each point depending on position: $w = 0$ if the pixel falls within 10 pc from the gap edge, increasing by 0.01 every 10 pc from the gap edge until $w=1$. The result is shown in the right-hand panel of Figure \ref{fig:combo}. Finally, we merged the combined KDE map with our full VMC surface density map, shown in the top panel of Figure \ref{fig:kde_map}. The resulting map has a mean stellar density of $3.21\times10^{-3}$ pc$^{-2}$, a median stellar density of $7.58\times 10^{-5}$ pc$^{-2}$ and a standard deviation of $7.65\times 10^{-3}$ pc$^{-2}$.

\section{Comparison with other surveys}
\label{appd:surveys}

\begin{figure*}
  \includegraphics[width=1\linewidth]{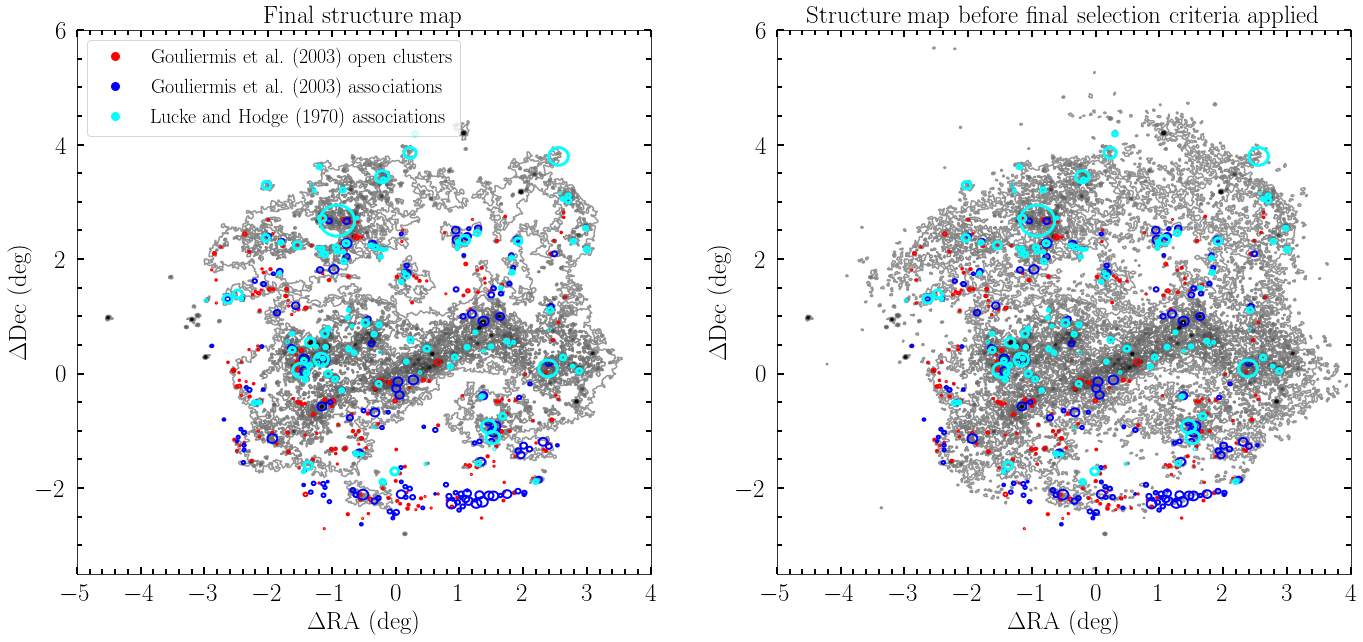}
  \caption{(Left) Our final structure map in greyscale with three surveys of LMC young stellar systems overplotted. (Right) Our structure map before the final selection criterion is applied, with the same three surveys overplotted. }
  \label{fig:overlap_map}
\end{figure*}

We compare our structures with published galaxy-wide catalogues of young stellar systems in the LMC. If the provided coordinates fall within the 1$\sigma$ isodensity contours, we estimate that we detected the system. If not, we estimate that system is not detected in our study.

The first survey used for this comparison is the \citet{lucke&hodge1970A} catalogue, containing 122 associations. They defined stellar associations as loose groups of bright blue stars, which they identified from $B$ and $V$ photographic plates. We detect 95\% of their associations in our study.

The next survey was published by \citet{gouliermis2003}. They identified young stellar systems from a photographic $U$ plate centred on the LMC. Their detection method is similar to ours: they identified contours of young systems on significance levels if they are 3$\sigma$ above the local mean stellar density. Then they classified the systems based on their calculated central densities. They provide two catalogues, one of their unbound and one of their intermediate stellar systems, which they describe as associations and open clusters, respectively. To compare with our structures, we impose that their stellar systems have at least five stars. After applying this criterion, their open cluster catalogue contains 231 systems and their association catalogue has 148 systems. We detect 77\% of the open cluster catalogue and 65\% of the association catalogue.

The final catalogue we compare with is the \citet{bica2008} catalogue, which is the most complete catalogue of stellar systems in the Magellanic Clouds. Their catalogue contains extended objects which were identified by eye on ESO/SERC $R$ and $J$ Sky Survey Schmidt films, as well as catalogues of systems available in the literature at the time of publication. We compare with objects classified as `CN' or `cluster in nebula', `NC' or `nebular with probable embedded cluster', and `NA' or `nebular with embedded association'. We chose these three classes because they were classified as the youngest stellar systems in the LMC by \citet{bonatto2010}, with reported ages of $\leq$10 Myr. There are 81 CN systems and we detect 95\%. There are 183 NC systems and we detect 93\%. There are 817 NA systems and we detect 88\%.

From the above analysis, it transpires that we have the best overlap with the \citet{bica2008} and \citet{lucke&hodge1970A} catalogues. The \citet{lucke&hodge1970A} and \citet{gouliermis2003} criteria are most similar to those used in our study, so we provide our structure map with the boundaries of their systems overplotted in Figure \ref{fig:overlap_map}. On the left we plot our final structure map, and on the right we plot our structure map before the final selection criterion is applied. The final selection criterion is that all of our structures on the 1$\sigma$ and 2$\sigma$ levels must contain at least one structure at the 3$\sigma$ and higher levels. This means that all of our structures on the 1$\sigma$ and 2$\sigma$ levels have at least 10 stars. Although \citet{gouliermis2003} used a very similar detection algorithm, the slight algorithm differences and differences in data caused them to find more structures in certain low-density regions. As Figure \ref{fig:overlap_map} shows, our algorithm detects most structures in the \citet{gouliermis2003} catalogue, but after we apply our final selection criteria we exclude $\geq$30 of them. We find that it is necessary to apply the final selection criterion, because we identify structures at levels above the median stellar density, which is effectively zero. In contrast, \citet{gouliermis2003} identified structures above the local mean. We must make a compromise to cull spurious objects and we made the decision to apply global selection criteria. Furthermore, the purpose of our study is to analyse the fractal properties, not to identify new stellar clusters or associations.

Finally, we estimated how many structures are newly identified objects by cross-matching our catalogue with the catalogues mentioned in this section. The central coordinates of our structures correspond to the geometric centroid. If our structure does not match within 0.5 arcsec the coordinates in one of those other surveys, we estimate that the structure of interest is a newly discovered stellar system. Doing this, we estimate that 45\% of our structures are newly discovered systems.

\bsp	
\label{lastpage}
\end{document}